\documentclass[10pt]{article}
\usepackage[footnotesize]{caption}
\usepackage{subcaption}
\usepackage{float}
\usepackage[symbol]{footmisc} 
\usepackage{graphicx}
\usepackage[superscript,sort]{cite}
\usepackage{array}
\usepackage{bm}
\usepackage{appendix}
\newcolumntype{x}[1]{%
>{\centering\hspace{0pt}}p{#1}}%
\usepackage[normalem]{ulem} 
\usepackage[colorlinks=true,linkcolor=black,citecolor=black,urlcolor=black,bookmarks=true,breaklinks=true]{hyperref}
\usepackage[usenames]{color}
\usepackage{amsmath,amssymb}

\usepackage{amsthm,amscd,amsxtra,amsfonts,amsmath,amssymb,multirow}
\usepackage{wrapfig}
\usepackage[tiny,compact]{titlesec}
\usepackage{threeparttable} 
\usepackage{helvet}

\usepackage{booktabs}
\usepackage[table]{xcolor}
\usepackage{xcolor,colortbl}
\usepackage{placeins}

\usepackage{tikz}
\usetikzlibrary{shapes,arrows}
\usepackage[T1]{fontenc}
\usepackage[linesnumbered,ruled]{algorithm2e} 
\titlespacing*{\section}{0pt}{*0}{*0}
\titlespacing*{\subsection}{0pt}{*0}{*0}
\titlespacing*{\subsubsection}{0pt}{*0}{*0} 
\titlespacing{\paragraph}{0pt}{*0}{*1}

\definecolor{MyPurple}{rgb}{1,0,1}

\renewcommand{\thesubsection}{{\thesection}.\Alph{subsection}}

\newcommand{\beq}[1]{\begin{equation} \label{#1}}
\newcommand{\eeq}{\end{equation}}
\newcommand{\barray}{\begin{array}{ll}}
\newcommand{\earray}{\end{array}}

\usepackage{longtable} 
\usepackage{threeparttable} 
\usepackage{booktabs} 

\begin{document}
\title{DG-GL: Differential geometry based geometric learning of molecular datasets }

\author{Duc Duy Nguyen$^{1}$ and  Guo-Wei Wei$^{1,2,3,}$\footnote{
		Corresponding to Guo-Wei Wei.		Email: wei@math.msu.edu}\\
$^1$ Department of Mathematics,
Michigan State University, MI 48824, USA.\\
$^2$ Department of Electrical and Computer Engineering,
Michigan State University, MI 48824, USA. \\
$^3$ Department of Biochemistry and Molecular Biology,
Michigan State University, MI 48824, USA. \\
}

\date{\today}
\maketitle

\begin{abstract}
\noindent {\bf Motivation:}
Despite its great success in various physical modeling, differential geometry (DG) has rarely been devised as a versatile tool for analyzing large, diverse and complex molecular and biomolecular datasets due to the limited understanding of its potential power in 
dimensionality reduction and its ability to encode essential chemical and biological information in differentiable manifolds. \\
{\bf Results:}
We put forward a differential geometry based geometric learning (DG-GL) hypothesis that the intrinsic physics of three-dimensional (3D) molecular structures lies on  a family of low-dimensional manifolds embedded in a high-dimensional data space.   
We encode crucial chemical, physical and biological information into  2D element interactive manifolds, extracted from a high-dimensional structural data space via a multiscale discrete-to-continuum mapping using  differentiable density estimators. Differential geometry apparatuses are utilized to construct element interactive curvatures  
in analytical forms for certain analytically differentiable density estimators. These low-dimensional differential geometry representations  are paired with a robust machine learning algorithm to showcase their descriptive and predictive powers  for large, diverse and complex molecular and biomolecular datasets. Extensive numerical experiments are carried out to demonstrated that the proposed DG-GL strategy outperforms other advanced methods in the predictions of drug discovery related protein-ligand binding affinity, drug toxicity, and molecular solvation free energy. \\
\end{abstract}


\pagestyle{empty}
 
\newpage
{\setcounter{tocdepth}{3} \tableofcontents  }

\newpage

\section{Introduction}

Geometric data analysis (GDA) of biomolecules concerns molecular structural representation,  molecular surface definition, surface meshing and volumetric meshing, molecular visualization, morphological analysis, surface annotation, pertinent feature extraction, et cetera at a variety of scales and dimensions  \cite{Corey:1953,Koltun:1965,Lee:1971,Richards:1977,Connolly85,Zap, Vorobjev:1997b,  Sanner:1996, Grant:2007, LLi:2013, LinWang:2015}.
Among them, surface modeling is a low-dimensional representation of biomolecules, an important concept in GDA \cite{kirby2000geometric}. Curvature analysis,  such as the smoothness and curvedness of a given biomolecular surface, is an important issue in molecular biophysics. For example, lipid spontaneous curvature and BAR domain mediated membrane curvature sensing are all known biophysical effects.  Curvature, as a measure how much a surface is deviated from being  flat \cite{Koenderink:1992}, is a major player in  molecular stereospecificity   \cite{Cipriano:2009}, the characterization of protein-protein and protein-nucleic acid interaction hot spots, and drug binding pockets  \cite{XFeng:2012a,XFeng:2013b,KLXia:2014a}, and the analysis of molecular solvation \cite{ Dzubiella:2006}. 

Curvature analysis is an important aspect of differential geometry (DG), which is a fundamental topic in mathematics and its study dates back to the 18th century. Modern differential geometry  encompasses a long list of branches or research topics and draws on differential calculus, integral calculus,  algebra and differential equation to study problems in geometry or differentiable manifolds. The study  of differential geometry is fueled by its great success in a wide variety of applications, from the curvature of space-time in Einstein's general theory of relativity,   differential forms  in electromagnetism \cite{deschamps1981electromagnetics},  to Laplace-Beltrami operator in cell membrane structures \cite{Wolfgang:2002, Soldea:2006}. How biomolecules assume complex structures and intricate shapes and why biomolecular complexes admit convoluted interfaces between different parts can also be described by differential geometry \cite{Wei:2009}. 

In  molecular biophysics, differential geometry of surfaces offers a natural tool to separate the solute from the solvent, so that the solute molecule can be described in a microscopic detail while the solvent is treated as a macroscopic continuum, rendering a dramatic reduction in the number of degrees of freedom. A differential geometry based multiscale   paradigm was proposed for large  biological systems, such as proteins,   ion channels, molecular motors, and viruses, which, in conjunction with their aqueous environment, pose a  challenge to both theoretical description and prediction due to a large number of degrees of freedom \cite{Wei:2009}. In 2005, the curvature-controlled geometric flow equations were introduced for molecular surface construction and solvation analysis \cite{Wei:2005}. In 2006, based on   the Laplace-Beltrami flow,  the first variational solvent-solute interface, the minimal molecular surface (MMS), was proposed  for molecular surface representation  \cite{Bates:2006,Bates:2008}.   Differential geometry based solvation models have been developed for solvation modeling   \cite{ZhanChen:2010a, ZhanChen:2010b, ZhanChen:2011a, ZhanChen:2012, DuanChen:2012a, DuanChen:2012b, Wei:2012, Daily:2013, Thomas:2013}. A family of  differential geometry based multiscale models has been used to couple implicit solvent models with molecular dynamics, elasticity and fluid flow \cite{Wei:2009,Wei:2012, Wei:2013,  DuanChen:2012a, DuanChen:2012b}.  Efficient geometric modeling strategies associated with differential geometry based multiscale models have been developed in both Lagrangian-Eulerian \cite{XFeng:2012a, XFeng:2013b} and Eulerian representations \cite{KLXia:2014a, mu2017geometric}.

Although the differential geometry based multiscale paradigm provides a dramatic reduction in dimensionality, quantitative analysis,  and useful predictions of solvation free energies \cite{ZhanChen:2012,BaoWang:2015a} and ion channel transport \cite{DuanChen:2012a, DuanChen:2012b, DuanChen:2013,Wei:2012,Wei:2013},  it works in the realm of physical models. Therefore, it has a relatively confined applicability and its performance depends on many factors, such as the implementation of the Poisson-Boltzmann equation  or the Poisson-Nernst-Planck, which in turn depends on the microscopic parametrization of atomic charges. Consequently, these  models have a limited representative power for  complex biomolecular structures and interactions.   

In addition to its use in biophysical modeling,  differential geometry has been devised for qualitative characterization of biomolecules  \cite{XFeng:2012a, XFeng:2013b}. In particular, minimum and maximum curvatures offer good indications of the concave and convex regions of biomolecular surfaces. This characterization was combined with surface electrostatic potential computed from the Poisson model to predict potential protein-ligand binding sites \cite{KLXia:2014a, mu2017geometric}.  Most recently,   the use of molecular curvature for quantitative analysis and the prediction of solvation free energies of small molecules have been explored \cite{DDNguyen:2016c}. However, the predictive power of this approach is limited due to the use of whole molecular curvatures. Essentially, chemical and biological information in the complex biomolecule is mostly neglected in this low-dimensional representation. 

Efficient representation of diverse small-molecules and complex macromolecules is of great  significance to chemistry, biology and material sciences.  In particular, this representation   is crucial for understanding protein folding, the interactions of protein-protein, protein-ligand, and protein-nucleic acid, drug virtual screening, molecular solvation, partition coefficient,  boiling point etc. Physically, these properties are generally known to be determined by a wide variety of non-covalent interactions, such as hydrogen bond, electrostatics, charge-dipole, induced dipole,  dipole-dipole, attractive dispersion, $\pi-\pi$ stacking, cation-$\pi$,   hydrophobicity, hydrophobicity,  and/or van der Waals interaction. However, it is impossible to accurately calculate these properties for diverse and complex molecules in massive datasets using rigorous quantum mechanics, molecular mechanics, statistical mechanics, and  electrodynamics.  
 
While differential geometry has the potential to provide an efficient representation of diverse molecules and complex biomolecules in large datasets, its current representative power is mainly hindered by the neglect of crucial chemical and biological information in the low-dimensional representations of high dimensional molecular and biomolecular structures and interactions.  One way  to retain chemical and biological information in differential geometry representation is to systematically break down a molecule or molecular complex into a family of fragments and then computing fragmentary differential geometry. Obviously, there is a wide variety of ways to create fragments from a molecule, rendering descriptions with controllable  dimensionality, and chemical and biological information. An element-level coarse-grained representation has been shown to be an appropriate choice in our earlier work \cite{BaoWang:2017FFTB, DDNguyen:2017d,ZXCang:2017c, ZXCang:2017b}. An important reason to  pursue element-level descriptions   is that the resulting representation needs to be scalable, namely,  being independent of the number of atoms in a given molecule so as to put molecules of different sizes in the dataset on an equal footing.   Additionally, fragments with specific element combinations can be used to  describe  certain types of non-covalent interactions, such as hydrogen bond,  hydrophobicity, and hydrophobicity that occur among certain types of elements. Most datasets provide either the atomic coordinates or three-dimensional (3D) profiles of molecules and biomolecules.  Mathematically, it is convenient to construct Riemannian manifolds on appropriately selected subsets of element types  to facilitate the use of differential geometry apparatuses. This manifold abstraction of complex molecular structures can be achieved via a discrete-to-continuum mapping in a multiscale manner  \cite{KLXia:2013d, KLXia:2015e, KLXia:2017}.  


The objective of the present work is to introduce  differential geometry  based geometric learning (DG-GL) as an accurate, efficient and robust representation  of molecular and biomolecular structures and their interactions. Our DG-GL assumption is that the intrinsic physics   lies on  a family of low-dimensional manifolds embedded in a high-dimensional data space.  The essential idea of our geometric learning  is to encipher crucial chemical,  biological and physical  information in the high-dimension data space into differentiable low-dimensional manifolds and then use differential geometry tools, such as Gauss map, Weingarten map, and fundamental forms,  to construct mathematical representations of the original dataset from the extracted manifolds. Using a multiscale discrete-to-continuum mapping, we introduce a family of Riemannian  manifolds,  called element interactive manifolds, to facilitate differential geometry analysis and compute element interactive curvatures.   The  low-dimensional differential geometry representation of high-dimensional molecular structures is paired with state-of-the-art machine learning algorithms to predict drug-discovery related molecular properties of interest, such as the free energies of solvation, protein-ligand binding  affinities, and drug toxicity. We demonstrate that the proposed  DG-GL strategy outperforms other cutting edge approaches in the field.

\section{Methods and  algorithms}\label{Methods}
This section describes methods and algorithms for geometric learning . We start by a review of a multiscale discrete-to-continuum mapping algorithm which extracts low-dimensional element interactive manifolds from high-dimensional molecular datasets.  Differential geometry apparatuses are applied to element interactive manifolds to construct appropriate mathematical representations suitable for machine learning, rendering  a DG-GL  strategy. 

\subsection{Element interactive manifolds}

\subsubsection{Multiscale discrete-to-continuum mapping} \label{Density}

Let ${\cal X}=\{{\bf r}_1, {\bf r}_2, \cdots, {\bf r}_N, \}$ be a finite set for $N$ atomic coordinates in a molecule and $q_j$ be the partial charge on the $j$th atom. Denote $\mathbf{r}_j \in \mathbb{R}^3$ the position of $j$th   atom, and $\|\mathbf{r} -\mathbf{r}_j\|$ the Euclidean distance between the $j$th  atom and a point ${\bf r} \in \mathbb{R}^3$. 
The   unnormalized molecular number density and molecular charge density are given by a discrete-to-continuum mapping \cite{KLXia:2013d,Opron:2014,DDNguyen:2016b}
 \begin{align}\label{fri_surface}
	\rho(\mathbf{r}, \{ \eta_j\}, \{w_j\})=\sum_{\substack{j=1}}^{N} w_j  \Phi\left(\|\mathbf{r}-\mathbf{r}_j\|;\eta_{j}\right),
\end{align}
where $w_j=1$ for molecular number density and $w_j=q_j$ for molecular charge density. Here, 
$ \eta_{j}$ are characteristic distances   and $\Phi$ is a $C^2$ correlation kernel or a density estimator  that satisfies the following admissibility conditions 
\begin{align}\label{eq:admiss}
\Phi \left(\|\mathbf{r}- \mathbf{r}_i\|;\eta_{j}\|\right)&=1, \quad{\rm as} \quad  \|\mathbf{r} -\mathbf{r}_j\| \rightarrow 0, \\
\Phi \left(\|\mathbf{r} - \mathbf{r}_j\|;\eta_{j}\|\right)&=0, \quad {\rm as} \quad  \|\mathbf{r} -\mathbf{r}_j\| \rightarrow \infty. 
\end{align}
Monotonically decaying  radial basis functions are all admissible. Commonly used  correlation kernels include  generalized exponential functions 
\begin{align}\label{exponential} 
\Phi\left(\|\mathbf{r} -\mathbf{r}_j\|;\eta_{ j}\|\right)=e^{-\left(\|\mathbf{r} -\mathbf{r}_j\|/\eta_{ j}\right)^\kappa}, \quad \kappa>0;
\end{align}  
and generalized Lorentz functions  
\begin{align}\label{Lorentz1}
\Phi\left(\|\mathbf{r} -\mathbf{r}_j\|;\eta_{ j}\right)=\frac{1}{1+\left(\|\mathbf{r} -\mathbf{r}_j\|/\eta_{ j}\right)^\nu},\quad \nu>0.
\end{align}
Many other functions, such as $C^2$ delta sequences of the positive type discussed in an earlier work \cite{GWei:2000} can be employed as well.

Note that $	\rho(\mathbf{r}, \{ \eta_j\}, \{w_j\})$ depends on  scale parameters $\{\eta_j\}$ and possible charges $\{q_j\}$. A multiscale representation can be obtained by choosing more than one set of  scale parameters.  It has been shown that molecular number density (\ref{fri_surface}) serves as an excellent representation of molecular surfaces \cite{KLXia:2015e}. 
However, differential geometry properties computed from $\rho(\mathbf{r}, \{ \eta_j\}, \{w_j\})$ 
have a very limited predictive power \cite{DDNguyen:2016c}.

\subsubsection{Element interactive densities }  \label{sec:EID}

Our goal is to develop a DG representation of molecular structures and interactions in large molecular or biomolecular datasets. More specifically, we are interested in the  description of  non-covalent intramolecular molecular interactions in a molecule and intermolecular interactions in molecular complexes, such as protein-protein, protein-ligand, and  protein-nucleic acid complexes. With large datasets in mind, we seek an efficient manifold reduction  of high dimensional structures. To this end, we extract common features in most molecules or molecular complexes.  In order to make our approach scalable, the structure of our descriptors must be uniform regardless of  the sizes of molecules or their complexes.   

We consider a systematical and  element-level description of molecular interactions. For example, in the protein-ligand interactions, we classify all interactions as those between  commonly occurring  element types in proteins and  commonly occurring  element types in  ligands. Specifically,  commonly occurring element types in proteins include ${\rm H, C, N, O, S}$ and commonly occurring element types in  ligands are ${\rm H, C, N, O, S, P, F, Cl, Br, I}$. Therefore, we have a total of 50 protein-ligand  element specific groups: ${\rm HH, HC, HO, \cdots, HI, CH, \cdots, SI}$. These 50 element-level descriptions  are devised as an approximation to non-covalent interactions in large protein-ligand binding datasets. 
In fact, due to the absence of ${\rm H}$ in most Protein Data Bank (PDB) datasets, we exclude hydrogen in protein element types. For this reason, we only consider a total of 40 element specific group descriptions of protein-ligand interactions in practice.
Similarly, we  have a total of 25 element specific group descriptions of protein-protein interactions while practically consider only 16  collective descriptions.  
This approach can be trivially extended to other interactive systems in chemistry, biology and material science. 

We denote the set of commonly occurring chemical element types in the dataset as  ${\cal C}=\{{\rm H, C, N, O, S, P, F, Cl,  \cdots }\}$. As such,  ${\cal C}_3= {\rm N} $ denotes the third  chemical element in the collection, i.e., a nitrogen element. The selection of   ${\cal C}$ is based on the statistics of the dataset. Certain rarely occurring chemical element types will be ignored in the present description.

 
For a molecule or molecular complex with $N$ commonly occurring atoms, its $i$th atom  is labeled both by its element type $\alpha_j$, its position ${\bf r}_j$ and partial charge $q_j$. The collection of these $N$ atoms is  set  ${\cal X} = \{(\mathbf{r}_j, \alpha_j,q_j)|\mathbf{r}_j\in {\rm I\!R}^3; \alpha_j \in {\cal C};  j=1,2,\ldots,N \}$. 

We assume that all the pairwise non-covalent interactions  between element types ${\cal C}_{k}$ and ${\cal C}_{k'}$ in a molecule or a molecular complex can be represented by correlation kernel $\Phi$
\begin{equation}\label{CollInter}
\{\Phi(||\mathbf{r}_i-\mathbf{r}_j||; \eta_{kk'})| \alpha_i  \in {\cal C}_{k}, \alpha_j  \in {\cal C}_{k'};  i,j = 1,2,\ldots,N; 
||\mathbf{r}_i-\mathbf{r}_j||> r_i+r_j +\sigma \},
\end{equation} 
 where $||\mathbf{r}_i-\mathbf{r}_j||$ is the Euclidean distance between the $i^{th}$ and $j^{th}$ atoms, $r_i $ and $ r_j$ are the atomic radii of $i^{th}$ and $j^{th}$ atoms, respectively and $\sigma$ is the mean value of the standard deviations of $r_i $ and $ r_k$  in the dataset. 
The distance constraint ($||\mathbf{r}_i-\mathbf{r}_j||> r_i+r_j +\sigma $) excludes covalent interactions. Here $\eta_{kk'}$ is a characteristic distance between the atoms, which depends only on their element types.   
 %
%


Let $B({\bf r}_i,r_i)$ be a ball with a center ${\bf r}_i$ and a radius $r_i$. The atomic-radius-parametrized van der Waals domain of all  atoms of $k$th element type $D_k:= \cup_{{\bf r}_i, \alpha_i \in {\cal C}_k } B({\bf r}_i, r_r)$.   
We are interested in the element interactive number density and  element interactive charge density due to all atoms of $k'$th element type at $D_k$ are given by                       
\begin{equation}\label{ESRI}
\rho_{kk'}({\bf r},  \eta_{kk'}) =\sum_{j }w_j \Phi(||\mathbf{r}-\mathbf{r}_j||;\eta_{kk'}), 
\quad {\bf r}  \in D_k, \alpha_j  \in {\cal C}_{k'}; ||\mathbf{r}_i-\mathbf{r}_j||> r_i+r_j +\sigma, \forall \alpha_i\in    {\cal C}_{k};    k\neq k',
\end{equation} 
where $w_j=1$ for element interactive number density and $w_j=q_j $ for element interactive charge density. 
    Moreover, when $k=k'$, each atom can contribute to both the van der Waals domain $D_k$ and the summary of the element interactive density.  Therefore, we define the element interactive number density and element interactive charge density  due to all atoms of $k$th element type at $D_k$ as
\begin{equation}\label{ESRI2}
\rho_{kk}({\bf r}, \eta_{kk}) = \sum_{j } w_j \Phi(||\mathbf{r}-\mathbf{r}_j||; \eta_{kk}), 
\quad {\bf r}  \in D^i_k,   \alpha_i  \in {\cal C}_{k};  \alpha_j  \in {\cal C}_{k}; ||\mathbf{r}_i-\mathbf{r}_j||> 2r_j +\sigma,
\end{equation} 
 where $D^i_k=B({\bf r}_i,r_i),  \alpha_i \in {\cal C}_k$ is the van der Waals domain of the $i$th atom of the $k$th element type. 
Obviously, element interactive density and element charge density are  collective quantities for a given pair of element types. It is a $C^{\infty}$  function defined on the domain enclosed by the boundary of  $D_k$  of the $k$th element type.  
 
Note that a family of element interactive manifolds is defined by varying a constant $c$
 \begin{equation}\label{manifold}
 \rho_{kk'}({\bf r},  \eta_{kk'})=c\rho_{\max}, \quad 0\leq c \leq 1 \quad {\rm and } \quad \rho_{\max}=\max\{\rho_{kk'}({\bf r},  \eta_{kk'})\}.
\end{equation} 
Figure \ref{fig:flowchart} illustrates a few element interactive manifolds. 

\subsection{Element interactive curvatures}

\subsubsection{Differential geometry of differentiable manifolds}

One aspect of differential geometry concerns the calculus defined on  differentiable manifolds.
Consider a $C^2$ immersion ${\bf f}: U\rightarrow {\mathbb R}^{n+1}
$, where $U\subset {\mathbb R}^n$ is an open set and $\overline{U}$
is compact \cite{Wolfgang:2002,Bates:2008,Wei:2009}. Here ${\bf f(u)} =(f_1({\bf
u}),f_2({\bf u}),\cdots,f_{n+1}({\bf u}))$ is a hypersurface element
(or a position vector), and ${\bf u}=(u_1,u_2,\cdots,u_n)\in U$.
Tangent vectors (or directional vectors) of ${\bf f}$ are
$X_i=\frac{\partial {\bf f}} {\partial u_i},i=1,2\cdots n$. The
Jacobi matrix of the mapping ${\bf f}$ is given by $D{\bf
f}=(X_1,X_2,\cdots,X_n)$. The first fundamental form is a symmetric,
positive definite metric tensor of ${\bf f}$, given by
$I(X_i,X_j):=(g_{ij})=(D{\bf f})^T\cdot(D{\bf f})$. Its matrix
elements can also be expressed as $g_{ij}=\left\langle
X_i,X_j\right\rangle$, where $\left\langle,\right\rangle$ is the
Euclidean inner product in ${\mathbb R}^n$, $i,j =1,2,\cdots,n$.

Let  ${\bf N}({\bf u})$ be the unit normal vector given by the Gauss
map ${\bf N}: U\rightarrow R^{n+1}$,
\begin{equation}\label{normal}
{\bf N}(u_1,u_2,\cdots,u_n):=X_1 \times X_2 \cdots \times X_n/ \|X_1
\times X_2 \cdots \times X_n\| \in \bot_{\bf u} {\bf f},
\end{equation}
where $``\times'' $ denotes the cross product. Here $\bot_{\bf
u}{\bf f}$ is the normal space of ${\bf f}$ at point ${\bf X}={\bf
f(u)}$, where the  position vector ${\bf X}$ differs much from
tangent vectors $X_i$. The normal vector ${\bf N}$ is perpendicular
to the tangent hyperplane $T_{\bf u} {\bf f}$ at ${\bf X}$. Note
that $T_{\bf u}{\bf f}\oplus \bot_{\bf u}{\bf f} =T_{{\bf
f(u)}}{\mathbb R}^n$, the tangent space at ${\bf X}$. By means of
the normal vector ${\bf N}$ and tangent vector $X_i$, the second
fundamental form is given by
\begin{equation}
II(X_i,X_j)=(h_{ij})_{i,j=1,2,\cdots
n}=\left(\left\langle-\frac{\partial{\bf N}}{\partial u_i},
X_j\right\rangle\right)_{ij}.
\end{equation}
The mean curvature can be calculated from
$ H=\frac{1}{n}h_{ij}g^{ji}, $
where we use the Einstein summation convention, and $(g^{ij})=(g_{ij})^{-1}$. The Gaussian curvature is given by $K=\frac{{\rm Det}(h_{ij})}{{\rm Det}(g_{ij})}$. 

\subsubsection{Element interactive curvatures}
Based on the above theory, the Gaussian  curvature ($K$) and the mean curvature ($H$)  of  element interactive density $\rho({\bf r})$ can be easily evaluated  \cite{Soldea:2006,KLXia:2014a}: 
	\begin{align}
	K=&\frac{1}{g^2}\left[
	2\rho_x \rho_y \rho_{xz}\rho_{yz} + 2\rho_x\rho_z\rho_{xy}\rho_{yz}+2\rho_y\rho_z\rho_{xy}\rho_{xz} \right. \nonumber\\
	&\left.  - 2 \rho_x \rho_z \rho_{xz} \rho_{yy} + 2 \rho_y \rho_z \rho_{xx} \rho_{yz} + 2 \rho_x \rho_y \rho_{xy} \rho_{zz} \right.\nonumber\\
	&\left. +\rho_z^2 \rho_{xx}  \rho_{yy} + \rho_x^2 \rho_{yy} \rho_{zz} + \rho_y^2 \rho_{xx} \rho_{zz}\right.\nonumber\\
	&\left. -\rho_x^2 \rho_{yz}^2 + \rho_y^2 \rho_{xz}^2 + \rho_z^2 \rho_{xy}^2 \right],
	\label{gaussian_curv}
	\end{align}
	and
	\begin{align}
	H=\frac{1}{2g^{\frac{3}{2}}}\left[
	2 \rho_x \rho_y \rho_{xy} + 2 \rho_x \rho_z \rho_{xz} + 2 \rho_y \rho_z \rho_{yz} - (\rho_y^2 + \rho_z^2)\rho_{xx} - (\rho_x^2 + \rho_z^2)\rho_{yy} - (\rho_x^2+\rho_y^2)\rho_{zz}\right],
	\label{mean_curv}
	\end{align}
	where $g=\rho_x^2 + \rho_y^2 + \rho_z^2$.
	With determined Gaussian and mean curvatures, the minimum curvature,  $\kappa_{\rm min}$, and maximum curvature, $\kappa_{\rm max}$, can be evaluated by
	\begin{align}
	\kappa_{\rm min}=\min\{H-\sqrt{H^2-K},H+\sqrt{H^2-K}\},\quad \kappa_{\rm max}=\max\{H-\sqrt{H^2-K},H+\sqrt{H^2-K}\}.\label{minmax_curv}
	\end{align}
	
Note that if we choose $\rho$ to be $\rho_{kk'}({\bf r}, \eta_{kk'})$ given in  Eq. (\ref{ESRI}), the associated element interactive curvatures (EIC) are continuous functions i.e.,  $K_{kk'}({\bf r}, \eta_{kk'}), H_{kk'}({\bf r}, \eta_{kk'}),  		\kappa_{kk',\rm min}({\bf r}, \eta_{kk'}), \kappa_{kk',\rm max}({\bf r}, \eta_{kk'}), \forall {\bf r}\in D_k$. These interactive curvature functions offer new descriptions of non-covalent interactions in molecules and molecular complexes.  In practical applications, we are particularly interested in evaluating EICs at the atomic centers and  define the element interactive Gaussian curvature (EIGC) by 
\begin{equation}\label{InterCurv}
   K^{\rm EI}_{kk'}(\eta_{kk'}) = \sum_i  K_{kk'}({\bf r}_i, \eta_{kk'}), \quad {\bf r}_i\in D_k; k\neq k' 
\end{equation}
and 
\begin{equation}\label{InterCurv2}
   K^{\rm EI}_{kk}(\eta_{kk}) =   \sum_i  K_{kk}({\bf r}_i, \eta_{kk}), \quad {\bf r}_i\in D^i_k, D^i_k \subset D_k. 
\end{equation}
Similarly, we can define   $ H^{\rm EI}_{kk'}(\eta_{kk'}),  		\kappa^{\rm EI}_{kk',\rm min}(\eta_{kk'})$ and $ \kappa^{\rm EI}_{kk',\rm max}( \eta_{kk'})$. In practical applications, these element interactive curvatures may involve a narrow band of manifolds.

Computationally, for 	interactive density densities based on correlation kernels  defined in Eqs. (\ref{exponential}) and  (\ref{Lorentz1}) their derivatives can be calculated analytically, and thus their EICs   can be evaluated analytically according to Eqs. \eqref{gaussian_curv}, \eqref{mean_curv} and \eqref{minmax_curv}.  The resulting analytical expressions are free of numerical error and directly suitable for molecular and biomolecular modeling.

\subsection{DG-GL strategy}
 
\begin{figure}[!ht]
	\centering
	 \includegraphics[width=\textwidth]{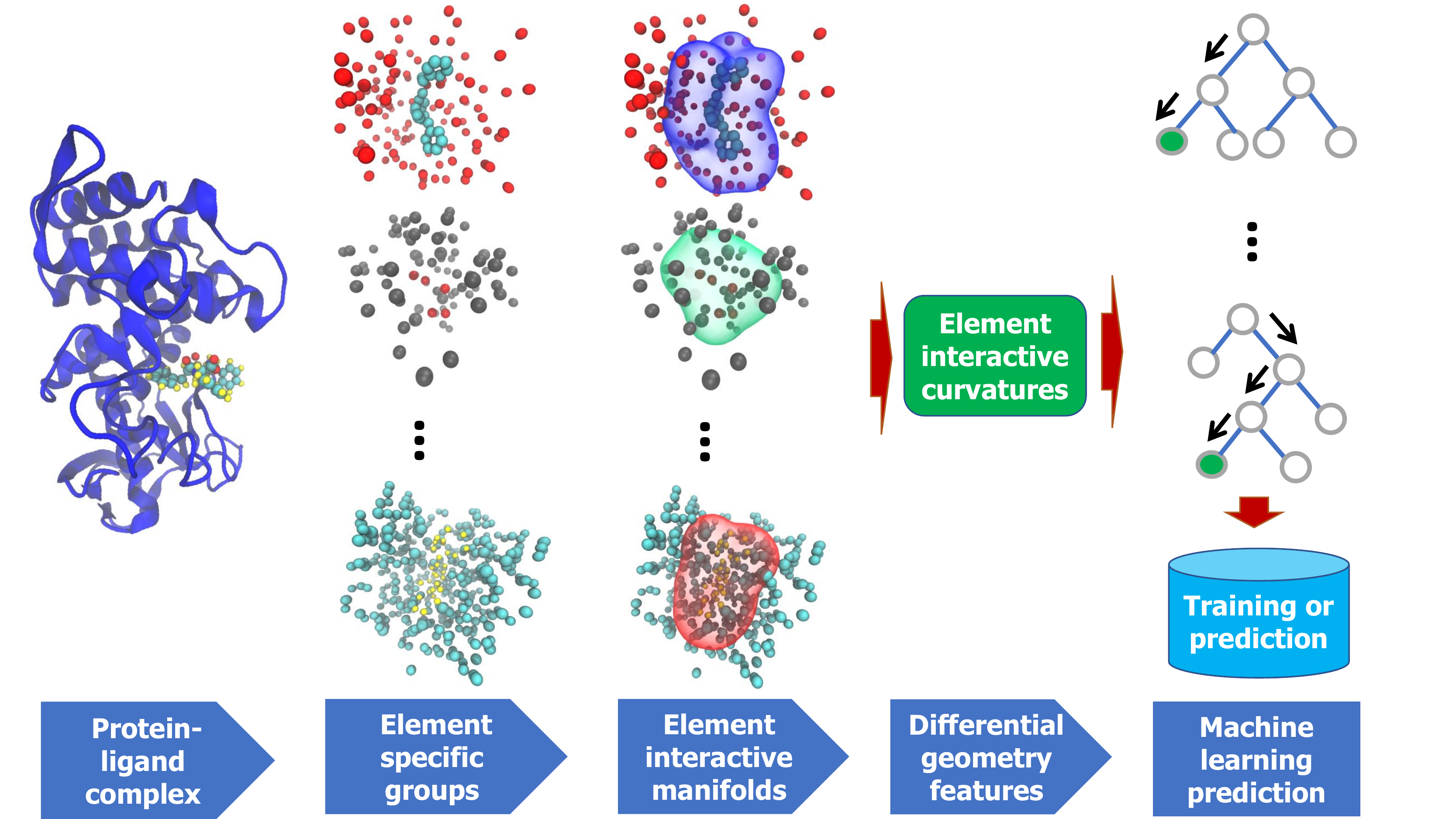} 
	\caption{ Illustration of  the DG-GL strategy using 1OS0 (first column).  In the second column, element specific groups are, from top to bottom, OC, NO, and CH, respectively.  Their corresponding element interactive manifolds are plotted in the third column, generated by setting the isovalue $c=0.01$. The differential geometry features (fourth column) are used in gradient boosting trees (last column) for training and prediction.}
	\label{fig:flowchart}
\end{figure}
 Paired with machine learning,  the proposed DG-GL of molecules is potentially highly powerful. In the training part of supervised learning (classification or regression), let ${\cal X}_i$  be the dataset from the $i$th molecule or molecular complex in the training dataset and $\mathbf{F}({\cal X}_i; \mathbf{\eta})$ is a function that maps the geometric information  into suitable DG representation with a set of parameters $\mathbf{\eta}$.     We cast the training into the following  minimization problem,
\begin{equation}
\min\limits_{\mathbf{\eta}, \mathbf{\theta}} \sum\limits_{i\in I}L(\mathbf{y}_i, \mathbf{F}({\cal X}_i; \mathbf{\eta}); \mathbf{\theta}),
\end{equation}
where $L$ is a scalar loss function to be minimized and  $\mathbf{y}_i$ is the collection of labels in the training set. Here $\mathbf{\theta}$ are the set of machine learning  parameters to be optimized and depend on machine learning algorithms chosen. Obviously, a wide variety of  machine learning algorithms, including random forest, gradient boosting trees, artificial  neural networks, and convolutional neural networks, can be employed in conjugation with the present DG representation. However, as our goal is to examine the representative  power of the proposed   geometric data analysis,  we only focus on the gradient boosting trees (GBTs) in the present work, instead of optimizing  machine learning algorithm selections.  Figure \ref{fig:flowchart} depicts the proposed DG-GL strategy. 

We use the  scikit-learn v0.19.1 package with the following parameters: n\_estimators=10000, max\_depth=7, min\_samples\_split=3, learning\_rate=0.01, loss=ls, subsample=0.3, max\_features=sqrt.  Our test indicates that random forest can yield similar results. Both ensemble methods are quite robust against overfitting \cite{ZXCang:2017a}.

\section{Results}

To  examine the validity,  demonstrate the utility, and illustrate the performance of the proposed DG-GL strategy for analyzing molecular and biomolecular datasets, we consider three representative problems. The first problem concerns  quantitative  toxicity prediction of small drug-like molecules. Quantitative toxicity analysis of new industrial products and new drugs has become a standard procedure required by the Environmental Protection Agency and the Food and Drug Administration. Computational analysis and prediction offer an efficient, relatively accurate, and low-cost approach for the toxicity virtual screening. There is always a demand for the next generation methods in toxicity analysis.  The second problem is about the solvation free energy prediction. Solvation is an elementary process in nature. Solvation analysis is particularly important for biological systems because water is abundant in living cells. The understanding of solvation is a  prerequisite for  the study of more complex chemical and biological processes in living organisms. The development of new strategies for the solvation free energy prediction is a major focus of molecular biophysics. 
In this work, we utilize toxicity and solvation to examine the accuracy and predictive power of the proposed DG-GL strategy for small  
molecular datasets. Finally, we consider protein-ligand bind affinity datasets to validate the proposed DG-GL strategy for analyzing biomolecules and their interactions with small molecules. Protein-ligand bind analysis is important for  drug design and discovery.   
The protocol of the proposed DG-GL strategy for solving these problems is illustrated Fig. \ref{fig:flowchart}.

\subsection{Model parametrization}
 
For the sake of convenience, we use notation ${\rm EIC}^{C}_{\alpha, \beta, \tau}$ to indicate the  element interactive curvatures (EICs)  generated by using curvature type $C$ with kernel type  $\alpha$ and corresponding kernel parameters $\beta$ and $\tau$. As such,  $C=K$, $C=H$, $C=k_{\min}$, and $C=k_{\max}$ represent Gaussian curvature, mean curvature, minimum curvature and maximum curvature, respectively. Here, $\alpha={\rm E}$ and $\alpha={\rm L}$ refer to generalized exponential and generalized Lorentz kernels, respectively. Additionally, $\beta$ is the kernel order such that $\beta=\kappa$ if $\alpha={\rm E}$, and $\beta=\nu$ if $\alpha={\rm L}$. Finally,  $\tau$ is used such that $\eta_{kk'}=\tau (\bar{r}_k + \bar{r}_{k'})$, where $\bar{r}_k$ and $\bar{r}_{k'}$ are the van der Waals radii of element type  $k$ and element type $k'$, respectively.  

We propose a  DG  representation in which multiple kernels are parametrized at different scale ($\eta$) values. In this work, we consider at most two kernels.  As a straightforward notation extension, two kernels can be parametrized  by  ${\rm EIC}^{C_1C_2}_{\alpha_1, \beta_1,\tau_1;\alpha_2,\beta_2,\tau_2}$.

\subsection{Datasets}
Three drug-discovery related  problems involving  small molecules and macro molecules and their complexes are considered in the present work to demonstrate the performance, validate the  strategy and  analyze the limitation of the proposed DG-GL strategy for molecular and biomelocular datasets. Details for these problems are described below.  

\subsubsection{Toxicity}
One of our interests is to examine the performance of our EIC on quantitative drug toxicity prediction. We consider an IGC$_{50}$ set which measures the concentration that inhibits the 50\% of the growth of  Tetrahymena pyriformis organism after 40 hours. This dataset was collected by Schultz and coworkers \cite{akers:1999, zhuhao:2008}. Its 2D SDF format molecular structures and toxicity end points (in log(mol/L) unit) are available on \href{https://www.epa.gov/chemical-research/toxicity-estimation-software-tool-test} {Toxicity Estimation Software Tool} (TEST) website. The 3D MOL2 molecular structures were created with the Schr\"{o}dinger software in our earlier work \cite{KDWu:2018a}.   The IGC$_{50}$ set consists of 1792 molecules that are split into a training set (1434 molecules) and a test set (358 molecules). The end point values lie between 0.334 log(mol/L) and 6.36 log(mol/L).

\subsubsection{Solvation}
We are also interested in exploring the proposed EIC method for solvation free energy prediction. A specific solvation dataset used in this work was  collected by Wang {\it et al} \cite{wangjm:2001} for the purpose of testing their  method named weighted solvent accessible surface area (WSAS). To validate our differential geometry approach, we consider the Model III in their work. In this model, a total of 387 neutral molecules in the 2D SDF format is divided into a training set (293 molecules) and  a test set (94 molecules)  \cite{wangjm:2001}. The 3D MOL2 molecular structures were created with the Schr\"{o}dinger software in our earlier work  \cite{BaoWang:2018FFTS}.   

\begin{table*}[!ht]
	\centering
	\caption{Summary  of PDBbind datasets used in the present work}
	\begin{tabular}{|l|c|c|c|}
		\hline
		& Total \# of complexes &  Train set complexes & Test set complexes\\ \hline
		PDBbind v2007  benchmark  & 1300 & 1105 & 195 \\\hline
		PDBbind v2013  benchmark  & 3711 & 3516 & 195 \\\hline
		PDBbind v2016  benchmark  & 4057 & 3767 & 290 \\
		\hline
	\end{tabular}
	\label{tab:PDBbind_datasets}
\end{table*} 

\subsubsection{Protein-ligand binding}
Finally, we are interested in using our EIC method to predict the binding affinities of protein-ligand complexes. A  standard benchmark for such a prediction is the PDBbind database \cite{RenxiaoWang:2009Compare, YLi:2014}. Three popular PDBbind datasets, namely PDDBind v2007, PDBbind v2013 and PDBbind v2016, are employed to test the performance of our method. Each PDBbind dataset has a hierarchical structure consisting of following subsets: a general set, a refined set, and a core set. The latter set is a subset of the previous one. Unlike other datasets used in this work, the PDBbind database provides 3D coordinates of ligands and their receptors obtained from experimental measurement via \href{https://www.rcsb.org/}{Protein Data Bank}. In each benchmark, it is standard to use the refined set, excluding the core set, as a training  set to build a predictive model for the binding affinities of the complexes in the test set (i.e., the core set). It is noted that the core set in the PDBbind v2013 is identical to that in PDBbind v2015. As a result, we use the  PDBbind v2015 refined set (excluding the core set) as the training set for the PDBbind v2013 benchmark. More information about these datasets is offered at the \href{http://PDBbind.org.cn/}{PDBbind website}. Table \ref{tab:PDBbind_datasets} lists the statistics of these three datasets used in the present study.

\subsection{Performance and discussion}

\subsubsection{Toxicity prediction}
Toxicity is the degree to which a chemical can damage an organism. These injurious events are called toxicity end points. Depending on the impacts on given targets, toxicity can be either quantitatively or qualitatively assessed. While the quantitative tasks report the minimal amount of chemical substances that can cause the fatal effects, the qualitative tasks classify chemicals into toxic and nontoxic categories. To verify the adverse response caused by chemicals on an organism, toxicity tests are traditionally conducted   {\sl in vivo} or {\sl in vitro}. However, such approaches usually reveal their shortcomings such as labor-intensive  and costly expense when dealing with a large number of chemical substances, not to mention the potential ethical issues. As a result, there  is a need to  develop efficiency computer-aided methods, or {\sl in silico} methods that are able to deliver an acceptable accuracy. There is a longstanding approach named quantitative structure activity relationship (QSAR). By assuming there is a correlation between structures and activities, QSAR methods can  predict the activities of new molecules without going through any real experiments in a wet  laboratory.

Many QSAR models have been reported in the literature in the past. Most of them are machine-learning based methods including a variety of traditional algorithms, namely regression and linear discriminant analysis \cite{deeb:2012}, nearest neighbors \cite{kauffman:2001, ajmani:2006}, support vector machine  \cite{deeb:2012,si:2007, du:2008} and random forest \cite{svetnik:2003}. In this toxicity prediction, we are interested in benchmarking our EIC method against other approaches presented in TEST \cite{test_guide} on the IGC$_{50}$ set.

As discussed in Section \ref{sec:EID}, we use 10 commonly occurring atom types,  ${\rm H, C, N, O, S, P, F, Cl, Br, I}$, for the element interactive curvature calculations, which results in 100 different pairwise combinations. Besides the use of element interactive curvatures, the statistical information, namely minimum, maximum, average and standard deviation, of the pairwise interactive curvatures as well as their absolute values is taken into account, which  leads to 800 additional features. In fact,  the atomic charge density is also used in the present work for generating  EICs of small molecules, which gives rise  to   a total number of 1800 features  for modeling the toxicity dataset.

To attain the best performance using EICs, the kernel parameters, i.e., $(\kappa, \tau)$ for exponential functions or $(\nu, \tau)$ for Lorentz functions, have to be optimized. To this end, we vary $\kappa$ or $\nu$ from 0.5 to 10 with an increment of 0.5, while $\tau$ values are chosen from 0.3 to 1.7 with an increment of 0.2. It is a common sense to use the cross-validation on the training data to obtain the optimal parameter set. For the toxicity dataset, we carry out a four-fold  cross-validation on the training set since we want each fold shares the similar size to the test set. Figures \ref{fig:tox-e-kfold} and \ref{fig:tox-l-kfold} report the optimal parameters of all different types of curvatures for exponential and Lorentz kernels, respectively. As shown in our previous work \cite{Opron:2015a, DDNguyen:2016b, Bramer:2018},  multiscale approaches can further boost the performance of one-scale models. Therefore, we add another kernel to the best single scale model selected from Figs. \ref{fig:tox-e-kfold} and \ref{fig:tox-l-kfold} to check if there is any improvement. As expected, multiscale models delivery  better cross-validation performances on the training data than their single-scale counterparts as shown in Figs. \ref{fig:tox-ee-kfold} and \ref{fig:tox-ll-kfold}. Specifically, while single-scale model using the mean curvature achieves the best R-squared correlation coefficient $R^2=0.743$ on the training set with parameters ${\rm EIC}^{H}_{E, 1.5, 0.3}$. The two-scale model ${\rm EIC}^{HH}_{E, 1.5, 0.3; E, 3.5, 0.3}$ produces a $R^2$ score as high as 0.772. In other words, the two-scale model learns the training data information more efficiency than its counterpart. 

\begin{table*}[!ht]
	\centering
	\caption{Comparison of  prediction results for the Tetrahymena Pyriformis IGC$_{50}$ test set.}
	\begin{tabular}{|c|c|c|c|c|c|c|}
		\hline
		Method & $R^2$ & $\frac{R^2-R_0^2}{R^2}$ & $k$ & RMSE & MAE & Coverage \\ \hline
		Hierarchical  \cite{test_guide}& 0.719 & 0.023 & 0.978 & 0.539 & 0.358 & 0.933 \\ 
		FDA  \cite{test_guide}& 0.747 & 0.056 & 0.988 & 0.489 & 0.337 & 0.978 \\ 
		Group contribution  \cite{test_guide}& 0.682 & 0.065 & 0.994 & 0.575 & 0.411 & 0.955 \\
		Nearest neighbor  \cite{test_guide}& 0.600 & 0.170 & 0.976 & 0.638 & 0.451 & 0.986 \\
		TEST consensus  \cite{test_guide} & 0.764 & 0.065 & 0.983 & 0.475 & 0.332 & 0.983 \\\hline
		\multicolumn{7}{|c|}{Results with EICs} \\ \hline
		EIC$^{k_{\min}}_{E, 10, 0.7}$  & 0.742 & 0.001 & 1.004 & 0.499 & 0.358 & 1.000 \\
		EIC$^{k_{\min}k_{\min}}_{E, 10, 0.7; E, 3.5, 0.3}$ & 0.767 & 0.003 & 1.002 & 0.477 & 0.338 & 1.000 \\
		EIC$^{k_{\min}}_{L, 5, 0.3}$   & 0.759 & 0.002 & 1.000 & 0.484 & 0.339 & 1.000 \\
		EIC$^{k_{\min}k_{\min}}_{L, 5, 0.3; L, 2, 1.3}$ & 0.767 & 0.002 & 1.002 & 0.476 & 0.329 & 1.000 \\
		Consensus$^{k_{\min}}$ & {\bf 0.781} & 0.004 & 1.003 & {\bf 0.463} & 0.324 & 1.000 \\
		\hline
		EIC$^{k_{\max}}_{E, 1, 0.3}$   & 0.749 & 0.001 & 0.999 & 0.492 & 0.344 & 1.00 \\
		EIC$^{k_{\max}k_{\max}}_{E, 1, 0.3; E, 3.5, 0.3}$ & 0.781 & 0.003 & 0.997 & 0.462 & 0.330 & 1.000 \\
		EIC$^{k_{\max}}_{L, 4, 0.5}$  & 0.748 & 0.001 & 0.998 & 0.494 & 0.352 & 1.000 \\
		EIC$^{k_{\max}k_{\max}}_{L, 4, 0.5; L, 4, 1.1}$ & 0.780 & 0.004 & 0.999 & 0.464 & 0.329 & 1.000 \\
		Consensus$^{k_{\max}}$    & 0.780 & 0.004 & 0.999 & 0.464 & 0.329 & 1.000 \\
		\hline
		EIC$^{K}_{E, 2.5, 0.3}$  & 0.725 & 0.001 & 1.001 & 0.516 & 0.366 & 1.000 \\
		EIC$^{KK}_{E, 2.5, 0.3; E, 1.5, 1.5}$ & 0.758 & 0.003 & 1.000 & 0.485 & 0.347 & 1.000 \\
		EIC$^{K}_{L, 2, 1.5}$  & 0.731 & 0.003 & 1.002 & 0.511 & 0.369 & 1.000 \\
		EIC$^{KK}_{L, 2, 1.5; L, 3, 0.3}$ & 0.769 & 0.005 & 1.001 & 0.476 & 0.342 & 1.000 \\
		Consensus$^K$    & {\bf 0.781} & 0.007 & 1.002 & 0.465 & 0.332 & 1.000 \\
		\hline
		EIC$^{H}_{E, 1.5, 0.3}$    & 0.745 & 0.001 & 1.000 & 0.497 & 0.349 & 1.000 \\
		EIC$^{HH}_{E, 1.5, 0.3; E, 3.5, 0.3}$   & 0.764 & 0.001 & 0.998 & 0.478 & 0.332 & 1.000 \\
		EIC$^{H}_{L, 5.5, 0.5}$    & 0.749 & 0.001 & 1.000 & 0.497 & 0.349 & 1.000 \\
		EIC$^{HH}_{L, 5.5, 0.5; L, 3.0, 1.3}$   & 0.773 & 0.003 & 1.000 & 0.471 & 0.325 & 1.000 \\
		Consensus$^H$      & 0.779 & 0.003 & 1.000 & 0.464 & {\bf 0.320} & 1.000 \\ \hline 
	\end{tabular}
	\label{tab:IGC50_results}
\end{table*}

After the training process, we are interested in seeing if a similar performance can be accomplished as one uses those models on the test set. The performance results of various types of curvatures on the IGC$_{50}$ test are reported in Table \ref{tab:IGC50_results}. Besides the R-squared correlation coefficient, we include the other common evaluation metrics, namely root-mean-squared error (RMSE) and mean-absolute error (MAE) for a general overview. To obtain predictions, we run gradient-boosting regressions up to 50 times for each model, then the final prediction is given by the average of these 50 predicted values. There are also consensus approaches presented in Table \ref{tab:IGC50_results}. The consensus models, named consensus$^C$, produce the average predicted values formed by the corresponding two-scale models EIC$^{C}_{E,\beta_1,\tau_1;E,\beta_2,\tau)2}$ and EIC$^{C}_{L,\beta_1,\tau_1;L,\beta_2,\tau_2}$. It is seen that consensus models consensus$^C$ typically offer a better performance than the rest of their counterparts.

To determine if a QSAR model has a predictive power, Golbraikh {\it et al} \cite{golbraikh:2003} proposed the following criteria
\begin{equation} \label{p0}
q^2 > 0.5, \quad R^2 > 0.6,  \quad  \frac{R^2-R_0^2}{R^2} < 0.1, \quad  {\rm and } \quad  0.85 \le k \le 1.15,
\end{equation}
where $q^2$ is the R-square correlation coefficient obtained by conducting the leave-one-out cross-validation (LOO CV) on the training set. $R^2$ is the squared Pearson correlation coefficient between experimental and predicted values of the test set. $R^2_0$, the R-square correlation coefficient between real and predicted values of the test set, is calculated by considering the linear regression without the intercept, and $k$ is the coefficient of that fitting line. It is easy to check that all our models reported in Table \ref{tab:IGC50_results} satisfy the last three evaluation criteria in \eqref{p0}. We do not carry on the LOO CV on the training data; therefore, the $q^2$ value is not available for this work. However, the 4-fold CV is conducted, and the $R^2$ values are always higher than 0.7 as shown in Figs. \ref{fig:tox-e-kfold}, \ref{fig:tox-ee-kfold}, \ref{fig:tox-l-kfold}, and \ref{fig:tox-ll-kfold}. LOO CV results would be typically better than those of the 4-fold CV. 
 
To illustrate the predictive power of the proposed EIC models, we present state-of-the-art results taken from the TEST software \cite{test_guide}  in Table \ref{tab:IGC50_results}. Since the approaches reported in Ref. \cite{test_guide} do not apply to the entire test data,  the coverage values of the TEST software are less than one.    Table \ref{tab:IGC50_results} confirms the state-of-the-art performances of various EIC models. All of our consensus models (Consensus$^X$) deliver a better prediction than the TEST consensus does, and the choice of the curvature type seems not affect the performances of our consensus models very much. Especially, the $R^2$ values of consensus$^{k_{\min}}$, consensus$^{k_{\max}}$, consensus$^K$, and Consensus$^H$ are found to be 0.781, 0.780, 0.781, and 0.779, respectively. In addition, the MAE  in log(mol/L) of the corresponding models are, respectively, as low as 0.324, 0.329, 0.332 and 0.320. These results are better than ones achieved by the TEST consensus \cite{test_guide} with its $R^2$ and MAE values being 0.764 and 0.332, respectively. As shown in Table \ref{tab:IGC50_results_testset} our best results were $R^2 =0.799$ and MAE$=0.315$, obtained by optimizing  kernel parameters according to the test set performance.  It is noted that in our earlier work using a combination of  both topological and physical features, the best prediction  has $R^2$ and MAE were 0.802 and 0.305 \cite{KDWu:2018a}. 
Since the mean curvature model  offers a balance between accuracy and variance among the different kernel selections, we will consider it as our primary model for the rest of our datasets.

\subsubsection{Solvation free energy prediction}
Solvation free energy is some of the most  important information in solvation analysis which can help to perceive other complex chemical and biological processes \cite{Daudel:1973,Kreevoy:1986,  Davis:1990a}. Therefore, it is essential to construct an accuracy scheme to predict  solvation free energies. In the past few decades, many theoretical methods have been reported in the literature for the solvation free energy prediction. Essentially, there are two types of physical models depending on the solvent molecules treatment, namely explicit and implicit ones. The typical explicit models refer to molecular mechanics  \cite{Martins:2014} and hybrid quantum mechanics/molecular mechanics  \cite{konig2014predicting} methods. In contrast, implicit models include many approaches, namely the generalized Born  model with various variants such as  GBSA \cite{Tan:2006c} and SM.x \cite{Truhlar:1999SolvationChemRev}, polarizable continuum model, and numerous derived forms of the Poisson-Boltzmann (PB) model \cite{Chung:2003, Sharp:1990, Honig:1995a, Gilson:1993, ZhanChen:2010a, ZhanChen:2011a,   BaoWang:2015a}. 

In this work, we are interested in examining our DG-GL strategy for solvation free energy predictions. To demonstrate the performance of the proposed model on relatively large datasets, we employ the solvation energy data set, Model III, collected by Wang {\it et al} \cite{wangjm:2001}. Model III has a total of 387 molecules (excluding ions) is split into a training data consisting of 293 molecules and test data with 94 molecules. To obtain a fair comparison, we use the same dataset splitting in our experiment except for omitting 4 molecules in the training set having the compound ID of 363, 364, 385 and 388 due to their obscure chemical names in the  PubChem database.  This omission results in a smaller training set of 284 molecules, which disfavors our method. Since we   deal with small molecules again, we employ the same feature generation procedure as that described  in the toxicity prediction.  

Since training data in the solvation free energy set differs from one in the toxicity task, one cannot expect to attain the optimal performance on the current set by reusing the kernel parameters found in the toxicity end points  prediction. To this end, we again carry out the parameter search by doing a 3-fold CV on the solvation training set. We use the mean absolute error as the main metric for such cross-validation, and only the mean curvature model is used in this task. Due to the observation of the toxicity set, we expect other curvature models will yield a similar performance. Figure \ref{fig:sol-el-kfold} depicts the performances of various kernel parameters from the 3-fold CV on the training set. Based on its heat-map plot, we conclude that EIC$^H_{{\rm E},3.5, 0.3}$ and EIC$^H_{{\rm L}, 3, 1.3}$ are the best single-kernel models. By using those kernel information, we naturally construct  two-kernel models and their performances on the training set are illustrated in Fig. \ref{fig:sol-eell-kfold}, which   reveals that EIC$^{HH}_{{\rm E},3.5, 0.3;{\rm E},2.5,1.3}$  and  EIC$^{HH}_{{\rm L}, 3, 1.3; {\rm L},6.5,0.3}$ are expected to be the best models for the test set prediction. After tuning  kernels' parameters, we use these models for solvation free energy prediction on the test set.

\begin{table*}[!ht]
	\centering
	\caption{Comparison of  prediction results for the solvation dataset collected by Wang {\it et al} \cite{wangjm:2001}.}
	\begin{tabular}{|c|c|c|c|}
		\hline
		Method & MAE (kcal/mol) & RMSE (kcal/mol) & $R^2$\\ \hline
		WSAS \cite{test_guide}& 0.66 & - & -\\ \hline
		FFT \cite{BaoWang:2018FFTS} & 0.57 & - & -\\
		\hline
		\multicolumn{4}{|c|}{Results with EICs} \\ \hline
		EIC$^H_{{\rm E},3.5, 0.3}$   & 0.575 & 0.921 & 0.904 \\
		EIC$^{HH}_{{\rm E},3.5, 0.3;{\rm E},2.5,1.3}$   & {\bf 0.558} & {\bf 0.857} & {\bf 0.920} \\
		EIC$^H_{{\rm L}, 3, 1.3}$    & 0.592 & 0.931 & 0.906 \\
		EIC$^{HH}_{{\rm L}, 3, 1.3; {\rm L},6.5,0.3}$    & 0.608 & 0.919 & 0.907  \\
		Consensus$^H$      & 0.567 & 0.862 & {\bf 0.920} \\ 
		\hline 
	\end{tabular}
	\label{solvation_results}
\end{table*}

To benchmark the proposed approach, we compare our results with the state-of-the-art methods, namely WASA \cite{wangjm:2001} and FFT \cite{BaoWang:2018FFTS}. The evaluations are reported in Table \ref{solvation_results}. Besides the MAE metric, we assess our models using additional ones such as root-mean-squared error (RMSE) and $R^2$; however, these metrics are missing from the literature. The results in Table \ref{solvation_results} indicate that our models, EIC$^{HH}_{{\rm E},3.5, 0.3;{\rm E},2.5,1.3}$  and Consensus$^H$, perform slightly better than the established ones in term of MAE. Specifically, our best model EIC$^{HH}_{{\rm E},3.5, 0.3;{\rm E},2.5,1.3}$ achieves MAE = 0.558 kcal/mol, while the WSAS and FFT attain MAE = 0.66 kcal/mol and MAE = 0.57 kcal/mol, respectively.  Unlike the toxicity prediction,  the consensus model in this experiment is not the best one.  However, if one does the blind prediction, the consensus is still the most reliable model. In fact, as shown in Table \ref{solvation_results_testset},  we obtained MAE = 0.524 kcal/mol and $R^2=0.931$ if we optimized our kernel parameters based the test set performance.

\subsubsection{Protein-ligand binding affinity prediction}
In order to demonstrate the application of our proposed element interactive curvature models on the various of biomolecule structures, we are interested in applying our EIC-score for the binding free energy prediction of a protein-ligand complex. There are numerous scoring functions (SFs) for the binding affinity estimation published in the literature. We can classify those SFs into four categories\cite{LiuJie:2014}: a) Force-field based or physical based scoring functions; b) Empirical or linear regression based scoring functions; c) Potential of the mean force (PMF) or knowledge-based scoring functions; and d) Machine learning based scoring functions. To validate the predictive power of the EIC-score, we employ three commonly used benchmarks, namely PDBbind v2007, PDBbind v2013, and PDBbind v2016 available online at \url{http://PDBbind.org.cn/}
  
To effectively capture the interactions between protein and ligand in a complex, we consider the scale factor $\tau$ and power parameters $\beta=\kappa \text{~or~} \nu$ in [0.5, 6] with an increment of 0.5. Moreover, we take the ideal low-pass filter (ILF) into account by considering   high $\beta$ values. To this end, we  assign $\beta$ $\in\{10, 15, 20\}$. The binding site of the complex is defined by a cut-off distance $d_c=20$\AA. The element interactive curvature is described by 4 commonly atom types, \{${\rm C, N, O, S}$\}, in protein and 10 commonly atom types, \{${\rm H, C, N, O, F, P, S, Cl, Br, I}$\}, in ligands. For a set of the atomic pairwise curvatures, one can extract 10 descriptive statistical values, namely sum, the sum of absolute values, minimum, the minimum of absolute values,  maximum, the  maximum of absolute values, mean, the mean of absolute values, standard deviation, and the standard deviation of absolute values, which results in a total of 400 features. Note that electrostatic curvatures are not employed in this study. 

Each benchmark involves its own training set; as a result, we design the different kernel parameters for the corresponding benchmark. We follow the same parameter search procedure as discussed  in the aforementioned datasets on toxicity and solvation predictions. Specifically, we carry out the 5-fold CV on  each training set with kernel parameters varying in their interested domains. Figures \ref{fig:v2007-el-kfold} and \ref{fig:v2015-el-kfold} plot the CV performance of single-kernel model on the training sets for PDBbind v2007 and PDBbind v2013 benchmarks. For one scale model, we found the exponential-kernel model and Lorentz-kernel model that produce the best Pearson correlation coefficient ($R_p$) for the PDBbind v2007 benchmark training set are, respectively, EIC$^{H}_{E,2,1}$ ($R_p=0.702$) and EIC$^{H}_{L,3.5,0.5}$ ($R_p=0.720$). While the optimal single-kernel models associated with exponential and Lorentz kernels for the PDBbind v2013 benchmark training set are EIC$^{H}_{E,1.5,5}$ ($R_p=0.754$) and EIC$^{H}_{L,5.5,5}$ ($R_p=0.758$). It is expected that a two-kernel model can boost the prediction accuracy; therefore, we again explore the utility of two-scale EIC-scores for binding affinity prediction. In the two-kernel models, the first kernel's parameters are fixed based on the previous finding. Then we search the parameters of the second kernel in the predefined space. Figures \ref{fig:v2007-eell-kfold} and \ref{fig:v2015-eell-kfold} depict the 5-fold CV performances of two-scale models on the PDBbind v2007 refined set and the PDBbind v2015 refined set, respectively. These experiments again confirm that multiscale models improve one-scale models' predictive power. Especially, the best choice of parameters and the mean values of $R_p$ form 5-fold CV on the PDBbind v2007 refined set for exponential-kernel  and Lorentz-kernel models are,  respectively, found to be EIC$^{HH}_{E,2,1;E,3,3}$ ($R_p=0.722$) and EIC$^{HH}_{L,3.5,0.5;L,3.5,2}$ ($R_p=0.729$). In addition, according to results in  \ref{fig:v2015-eell-kfold}, the best models for the PDBbind v2015 refined set for exponential-kernel and Lorentz-kernel models are, respectively, found to be  EIC$^{HH}_{E,1.5,5;E,3.5,3}$ ($R_p=0.771$) and EIC$^{HH}_{L,4.5,2.5;L,5.5,5}$ ($R_p=0.772$).

\begin{table*}[!ht]
	\centering
	\caption{Predictive performance of various models on the PDBbind v2007 benchmark.}
	\begin{tabular}{|c|c|c|}
		\hline
		Method & $R_p$ & RMSE (kcal/mol) \\
		\hline
		EIC$^{H}_{{\rm E},2,1}$    & 0.802 & 2.069 \\
		EIC$^{HH}_{{\rm E},2,1;{\rm E},3,3}$   & 0.812 & 1.999 \\
		EIC$^{H}_{{\rm L},3.5,0.5}$  & 0.778 & 2.131 \\
		EIC$^{HH}_{{\rm L},3.5,0.5;{\rm L},3.5,2}$   & 0.802 & 2.024  \\
		Consensus$^H$     & {\bf 0.817} & {\bf 1.987} \\ 
		\hline 
	\end{tabular}
	\label{tab:PDBbindv2007_results}
\end{table*}

\begin{figure}[!ht]
	\includegraphics[width=0.7\textwidth]{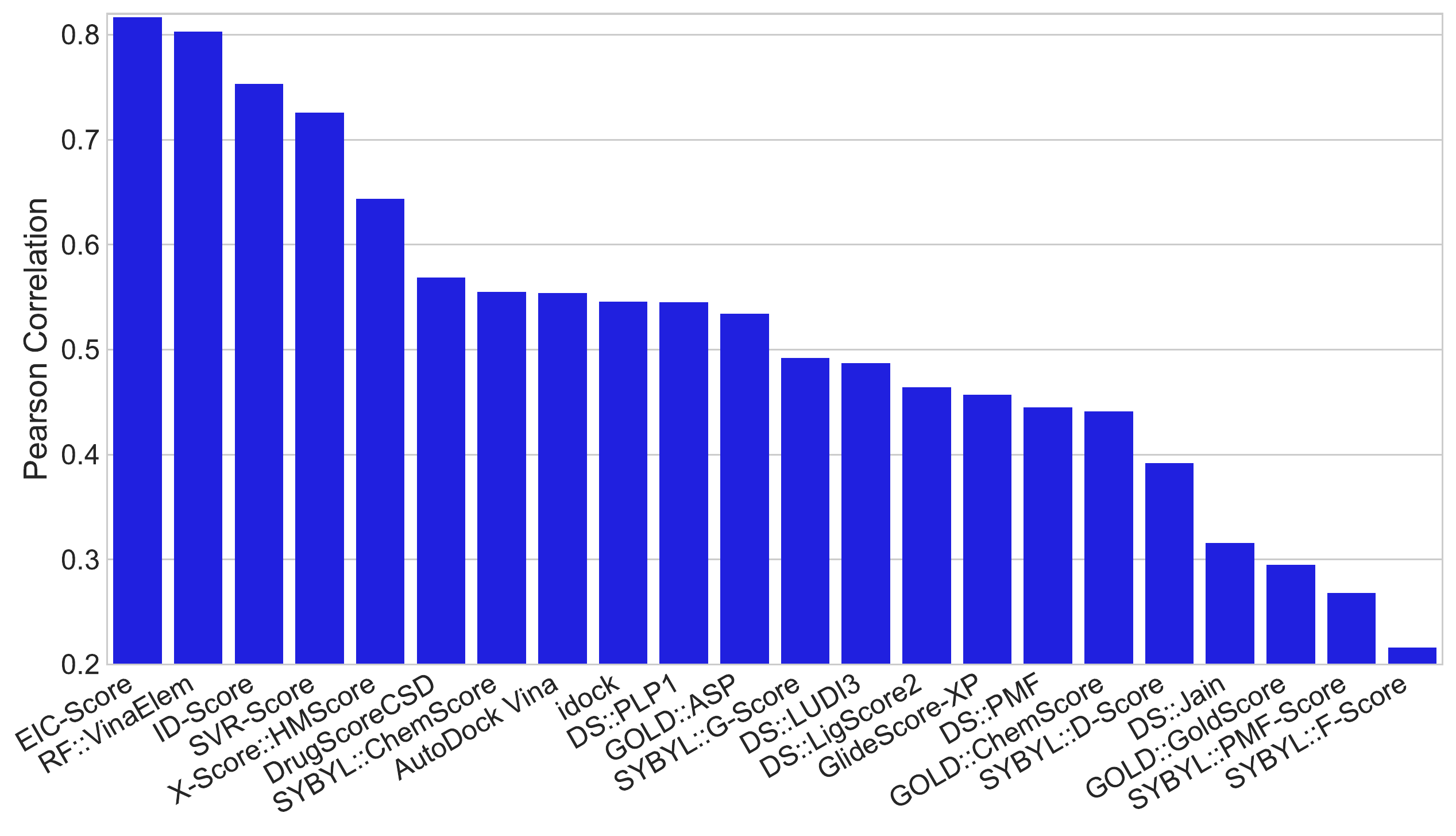}
	\caption{Performance comparison different scoring functions on the PDBbind v2007 core set. The Pearson correlation coefficients of other methods are taken from Refs. \cite{RenxiaoWang:2009Compare,Pedro:2010Binding, IDScore:2013,istar:2014,HLi:2015}. The proposed DG-GL strategy based scoring function, EIC-Score, achieves $R_p=0.817$ and RMSE=1.987 kcal/mol.}
	\label{fig:v2007-benchmark}
\end{figure}

\begin{table*}[!ht]
	\centering
	\caption{Predictive performance of various models on the PDBbind v2013 benchmark.}
	\begin{tabular}{|c|c|c|}
		\hline
		Method & $R_p$ & RMSE (kcal/mol) \\
		\hline
		EIC$^{H}_{E,1.5,5}$     & 0.755 & 2.060 \\
		EIC$^{HH}_{E,1.5,5;E,3.5,3}$   & 0.766 & 2.045 \\
		EIC$^{H}_{L,5.5,5}$   & 0.754 & 2.073 \\
		EIC$^{HH}_{L,4.5,2.5;L,5.5,5}$   & 0.770 & 2.032  \\
		Consensus$^H$      & {\bf 0.774} & {\bf 2.027} \\ 	
		\hline 
	\end{tabular}
	\label{tab:PDBbindv2015_results}
\end{table*}

\begin{figure}[!ht]
	\includegraphics[width=0.7\textwidth]{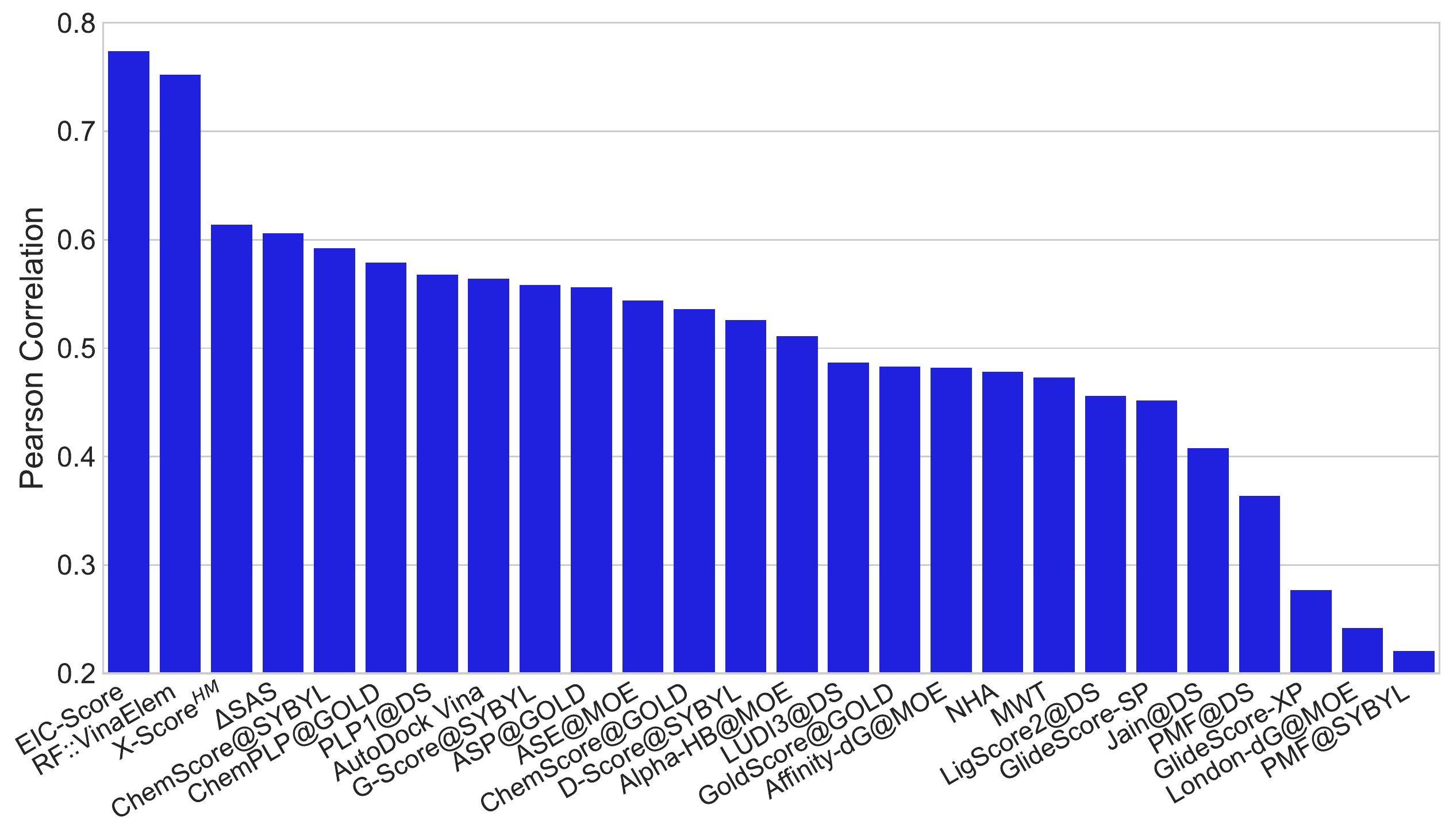}
	\caption{Performance comparison different scoring functions on the PDBbind v2013 core set. {{The performances of RF::VinaElem is adopted from Ref. \cite{HongJianLi:2015}. Results of 20 other scoring functions were reported in Ref. \cite{YLi:2014}}}. The proposed geometric  learning strategy based scoring function, EIC-Score, achieves $R_p=0.774$ and RMSE=2.027 kcal/mol.}
	\label{fig:v2013-benchmark}
\end{figure}

\begin{table*}[!ht]
	\centering
	\caption{Predictive performance of various models on the PDBbind v2016 benchmark.}
	\begin{tabular}{|c|c|c|}
		\hline
		Method & $R_p$ & RMSE (kcal/mol) \\
		\hline
		EIC$^{H}_{E,1.5,5}$   &{\bf 0.828} & {\bf 1.750} \\
		EIC$^{HH}_{E,1.5,5;E,3.5,3}$    & 0.825 & 1.762 \\
		EIC$^{H}_{L,5.5,5}$     & 0.809 & 1.816 \\
		EIC$^{HH}_{L,4.5,2.5;L,5.5,5}$   & 0.815 & 1.797  \\
		Consensus$^H$      & 0.825 & 1.767 \\ 
		\hline 
	\end{tabular}
	\label{tab:PDBbindv2016_results}
\end{table*}

\begin{figure}
	\centering
	\includegraphics[width=0.5\textwidth]{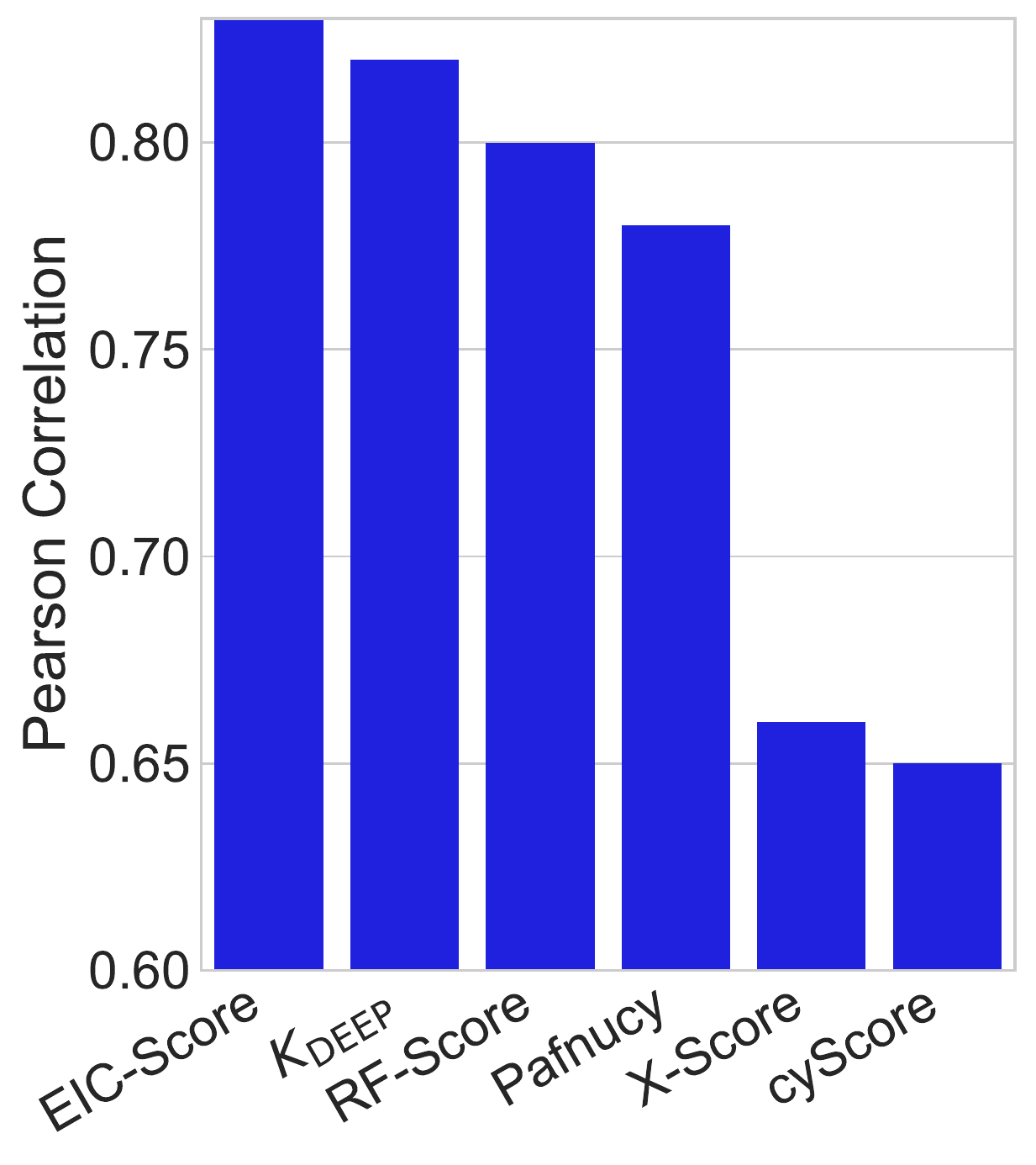}
	\caption{Performance comparison of different scoring functions on the PDBbind v2016 core set.  While the performances of $K_{\rm DEEP}$, RF-Score, X-Score, and cyScore are adopted from Ref. \cite{jimenez2018k}, Pearson correlation coefficient of Pafnucy is reported in Ref. \cite{2017arXiv171207042S}. The proposed DG-GL strategy based scoring function, EIC-Score, achieves $R_p=0.825$ and RMSE=1.767 kcal/mol.}
	\label{fig:v2016-benchmark}
\end{figure}

Having  validated EIC models, we are interested in applying them for the test set predictions  to see if they comply with their CV accuracies on the training sets. Table \ref{tab:PDBbindv2007_results} lists the accuracies in term of $R_p$ and RMSE for the test set of 195 complexes in the PDBbind v2007 benchmark. As expected the multiscale models outperform the single-scale counterparts. The best performance is achieved by the consensus of two-scale models (Consensus$^H$). Its $R_p$ and RMSE values are reported as 0.817 and 1.987 kcal/mol, respectively. In addition, we compare the predictive power of our EIC-Score to different scoring functions taken from Refs. \cite{RenxiaoWang:2009Compare,Pedro:2010Binding, IDScore:2013,istar:2014,HLi:2015} by plotting all of them in Fig. \ref{fig:v2007-benchmark}. Clearly, our model outperforms all the other scoring functions in this benchmark. In the PDBbind v2013 benchmark, we employ the kernel parameters optimized for the training data of this benchmark. Table \ref{tab:PDBbindv2015_results} reports the performance of our various EIC models on the test set of the PDBbind v2013 benchmark consisting of 195 complexes. Again, the two-kernel models outperform the one-kernel model, and the consensus approach delivers the best performance. Model Consensus$^H$ achieves $R_p = 0.774$ and RMSE = 2.027 kcal/mol. In this benchmark, we also compare our EIC-Score to various scoring functions in which results for 20 models are adopted from Ref. \cite{YLi:2014} and RF::VinaElem is reported in Ref. \cite{HongJianLi:2015}. Impressively, our EIC-Score model again stands out from the state-of-the-art scoring functions. All of the $R_p$ values of different models are plotted in Fig. \ref{fig:v2013-benchmark}, which confirms the utility of our model on the diversified protein-ligand binding datasets. In the last benchmark of the protein-ligand binding prediction task, we study the accuracy of our EIC-Score on the PDBbind v2016 benchmark. Since the PDBbind v2016 is a newer version of PDBbind v2015 with a supplement of a few recent complexes, we reuse the kernel parameters which are already optimized for the PDBbind v2015 training set. Table \ref{tab:PDBbindv2016_results} lists the $R_p$ and RMSE values of various EIC models. Surprisingly, the single-scale model 	EIC$^{H}_{E,1.5,5}$ is the best scoring function with $R_p=0.828$ and RMSE=1.750 kcal/mol. However, the consensus model 	Consensus$^H$  is very close behind with $R_p=0.825$ and RMSE=1.767. Since the PDBbind v2016 benchmark is relatively recent,  only a small number of models has been tested on thos benchmark.  Figure \ref{fig:v2016-benchmark} plots the performances of our EIC-Score along with other scoring functions reported in the literature. Especially, while $K_{\rm DEEP}$, RF-Score, X-Score, and CyScore are adopted from Ref. \cite{jimenez2018k}, Pafnucy model is taken from Ref. \cite{2017arXiv171207042S}. In this benchmark, our model is still the best performer which rigorously affirms the promising applications of our EIC-Score in the drug virtual screening and discovery.

In fact, as shown in Tables  \ref{tab:PDBbindv2007_results_testset}, \ref{tab:PDBbindv2015_results_testset} and 
\ref{tab:PDBbindv2016_results_testset},  if we optimized our kernel parameters based the test set performance, we obtained  $R=$0.825, 0.798, and 0.834 for predicting PDBbind v2007 core set, v2013 core set and v2016 core set, respectively, with associated  RMSE= 1.967,  1.52, and 1.746 kcal/mol, respectively , 

\section{Conclusion}
Differential geometry concerns the geometric structures on differentiable manifolds and  has been widely applied to  the general theory of relativity, differential forms in electromagnetism, and  Laplace-Beltrami flows in molecular and cellular biophysics. However,  differential geometry is rarely used in molecular and biomolecular data analysis.  Our earlier work indicates that although molecular manifolds and associated geometric properties  are able to provide a low-dimensional  description of  molecules and biomolecules, they have a very limited predictive power for large molecular datasets \cite{DDNguyen:2016c}. In particular, the potential role of differential geometry for drug design and discovery is essentially unknown. This work introduces differential geometry based geometric learning (DG-GL) as an  accurate, efficient and robust strategy for analyzing large, diverse and complex molecular and biomolecular datasets.  Based on the hypothesis that the most important physical and biological properties of molecular datasets still lie on an ensemble of low dimensional manifolds embedded in a high-dimensional data space, the key for success is how to effectively encode essential  chemical physical and biological information into low-dimensional manifolds. To this end, we propose element interactive manifolds,  extracted from the high-dimensional data space  via a multiscale discrete-to-continuum mapping, to  enable the embedding of  crucial chemical and biological information.   Differential geometry representations of complex molecular structures and interactions in terms of geodesic distances, curvatures,  curvature tensors etc are constructed from element interactive manifolds. The resulting geometric data analysis  is integrated with machine learning to predict various chemical and physical properties from large, diverse and complex molecular and biomolecular datasets. Extensive numerical experiments indicate that the proposed DG-GL strategy is able to outperform other state-of-the-art methods in drug toxicity, molecular solvation, and protein-ligand binding affinity predictions.

\section*{Acknowledgments}
 This work was supported in part by NSF grants  DMS-1721024, DMS-1761320  and IIS-1302285,  and MSU Center for Mathematical Molecular Biosciences Initiative.

\clearpage
\section*{Literature cited}
\renewcommand\refname{}

\bibliographystyle{unsrt}

\clearpage
 \appendix

\section*{Supplementary material }
\renewcommand{\thesubsection}{\Alph{subsection}}

\subsection{Toxicity prediction}
 
\subsubsection{Parameter search using the training set cross-validation}
\begin{figure}[!ht]
	\centering
	\begin{subfigure}{0.4\textwidth} 
		\includegraphics[width=\textwidth]{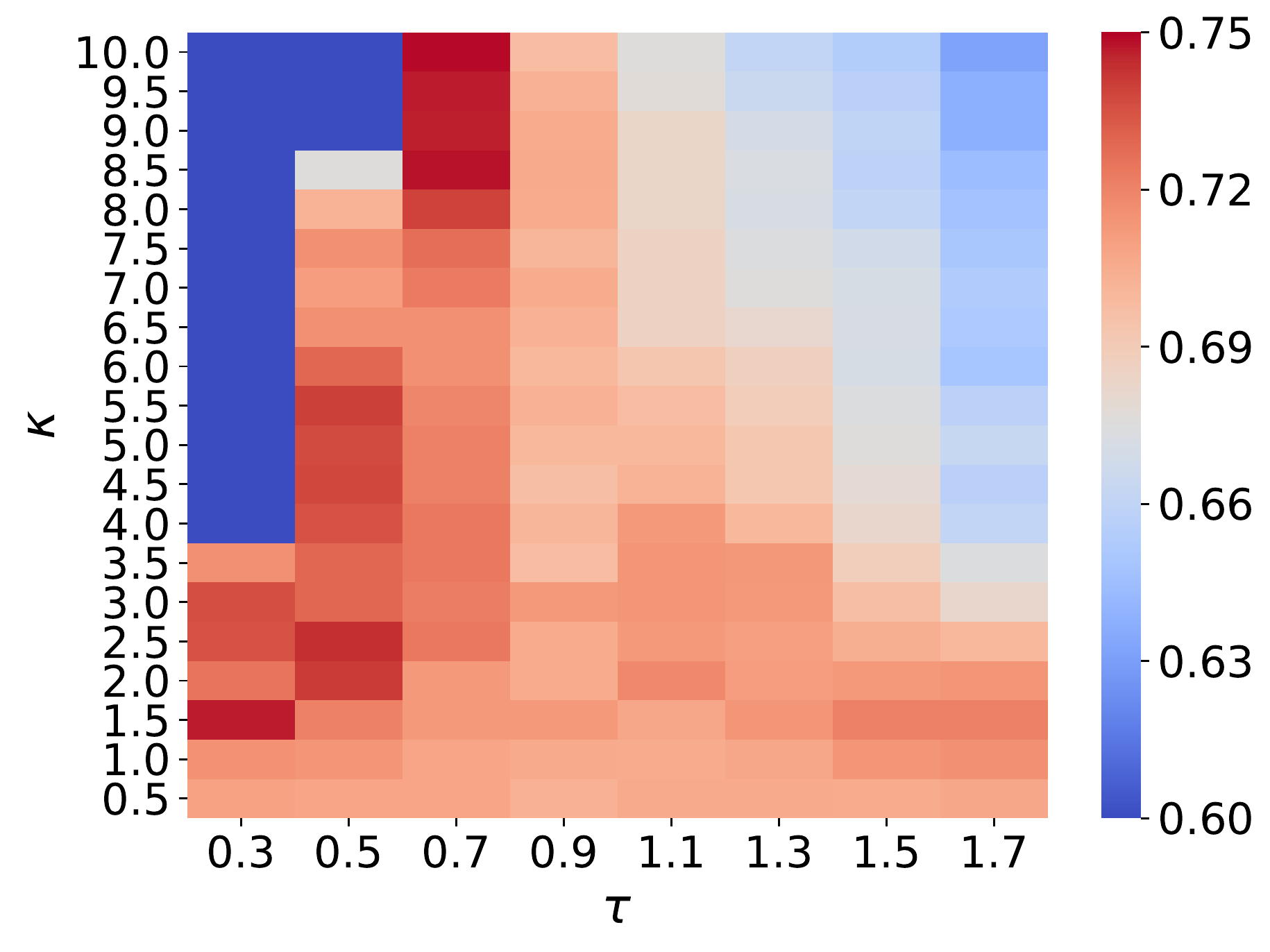}
		\caption{Minimum curvature} 
	\end{subfigure}
	\vspace{1em} 
	\begin{subfigure}{0.4\textwidth} 
		\includegraphics[width=\textwidth]{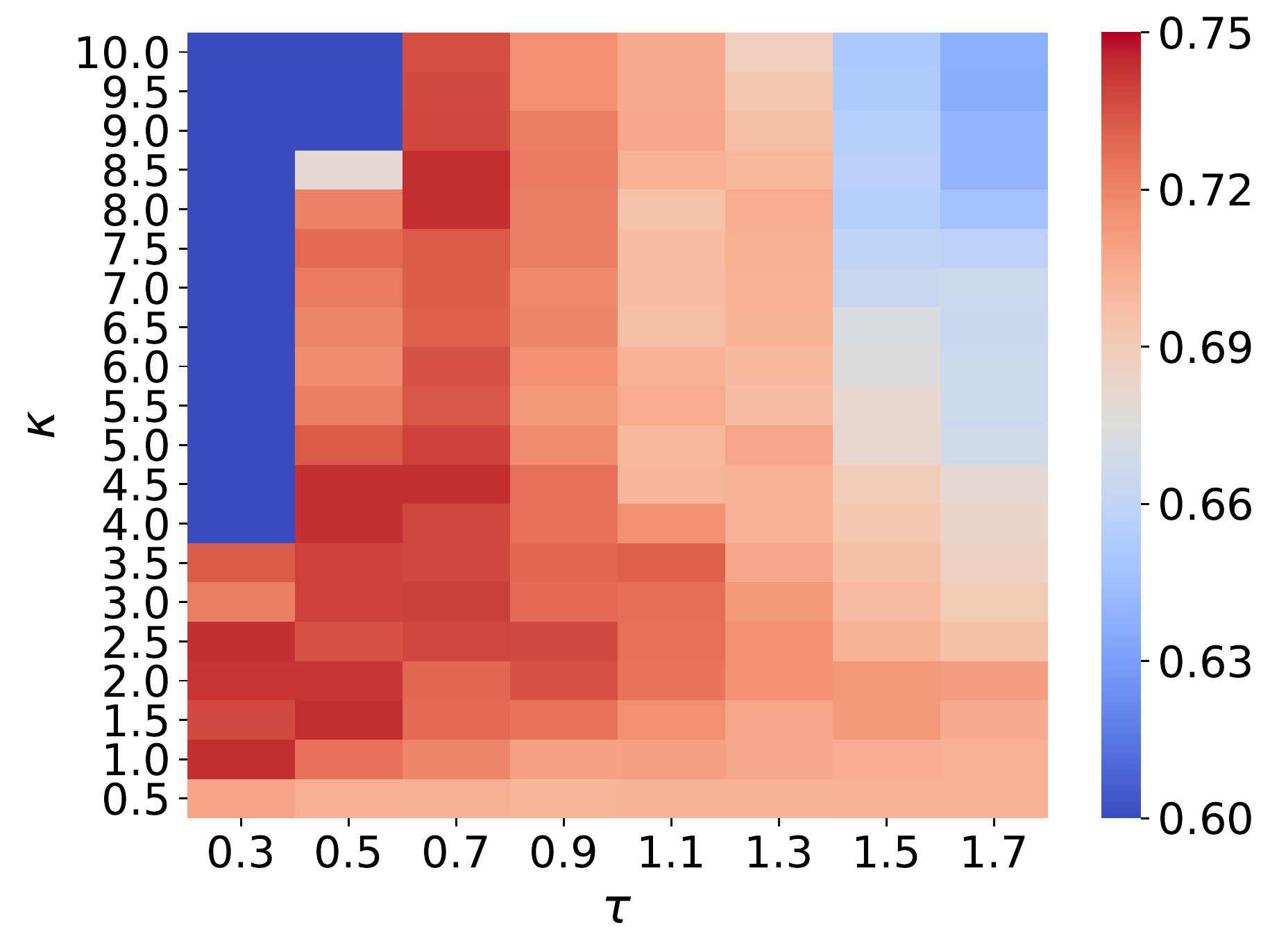}
		\caption{Maximum curvature} 
	\end{subfigure}
	\\
	\centering
	\begin{subfigure}{0.4\textwidth} 
		\includegraphics[width=\textwidth]{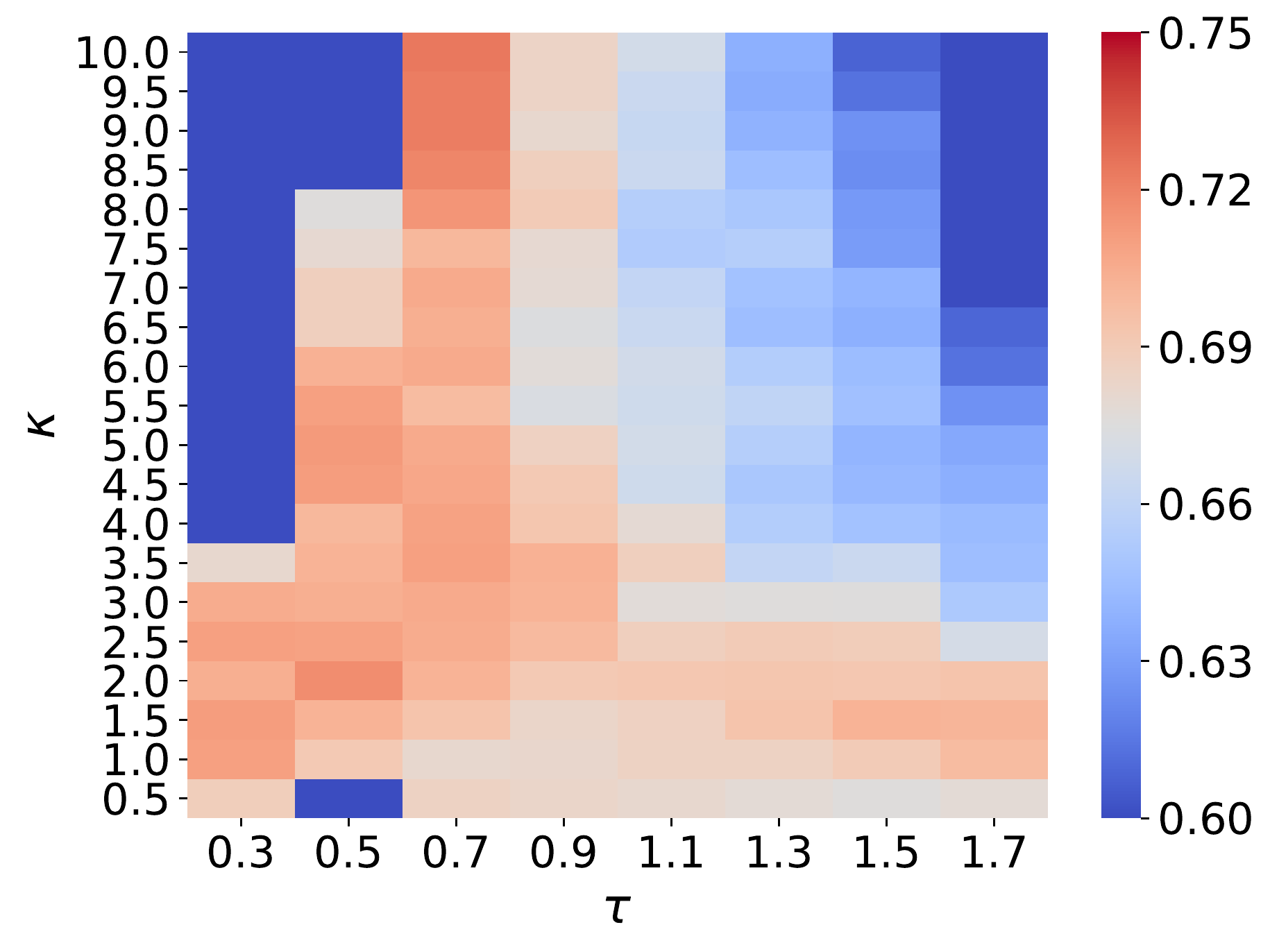}
		\caption{Gaussian curvature} 
	\end{subfigure}
	\vspace{1em} 
	\begin{subfigure}{0.4\textwidth} 
		\includegraphics[width=\textwidth]{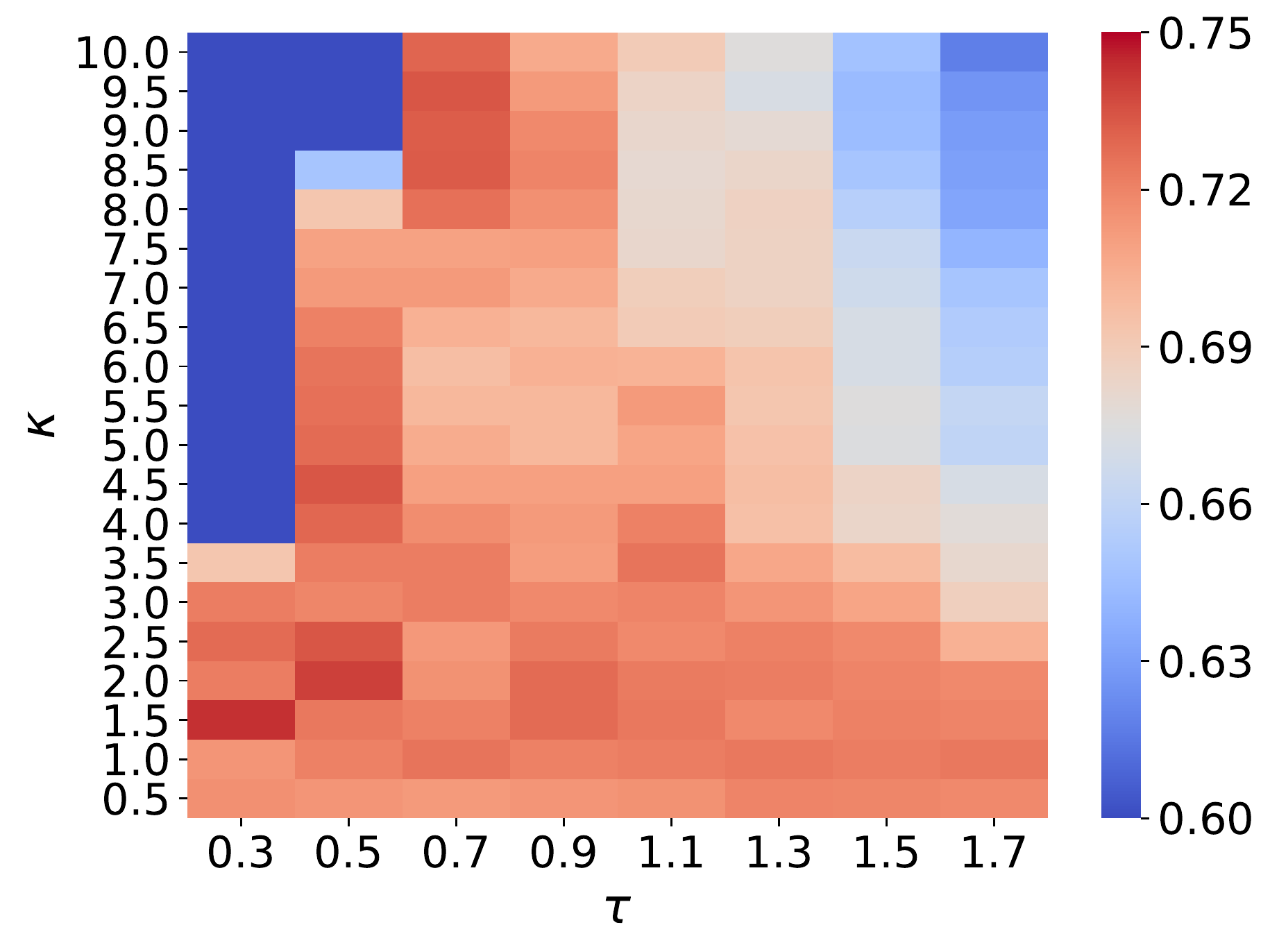}
		\caption{Mean curvature} 
	\end{subfigure}
	\caption{Median values of R-squared correlations $(R^2)$ from 4-fold cross validation performances of ${\rm EIC}^{X}_{{\rm E}, \kappa, \tau}$ on the IGC$_{50}$ training set are plotted against different values of $\tau$ and $\kappa$. Exponential kernels are utilized for curvature features generation.
		The best performance for different kinds of curvatures is found as follows 
		(a) minimum curvature: $(\tau=0.7, \kappa=10)$ with $R^2= 0.749$;
		(b) maximum curvature: $(\tau=0.3, \kappa=1)$ with $R^2= 0.744$;
		(c) Gaussian curvature: $(\tau=0.7, \kappa=10)$ with $R^2= 0.724$;
		(d) mean curvature $(\tau=0.3, \kappa=1.5)$ with $R^2= 0.743$.}
	\label{fig:tox-e-kfold}
\end{figure}

We here use 4-fold cross-validation on the training dataset to select the best parameters. Figure \ref{fig:tox-e-kfold} illustrates the 4-fold cross-validation (CV) performance of ${\rm EIC}^{C}_{{\rm E}, \kappa,\tau}$ on the IGC$_{50}$ training set against the different choices of $\kappa$ and $\tau$.

\begin{figure}[!ht]
	\centering
	\begin{subfigure}{0.4\textwidth} 
		\includegraphics[width=\textwidth]{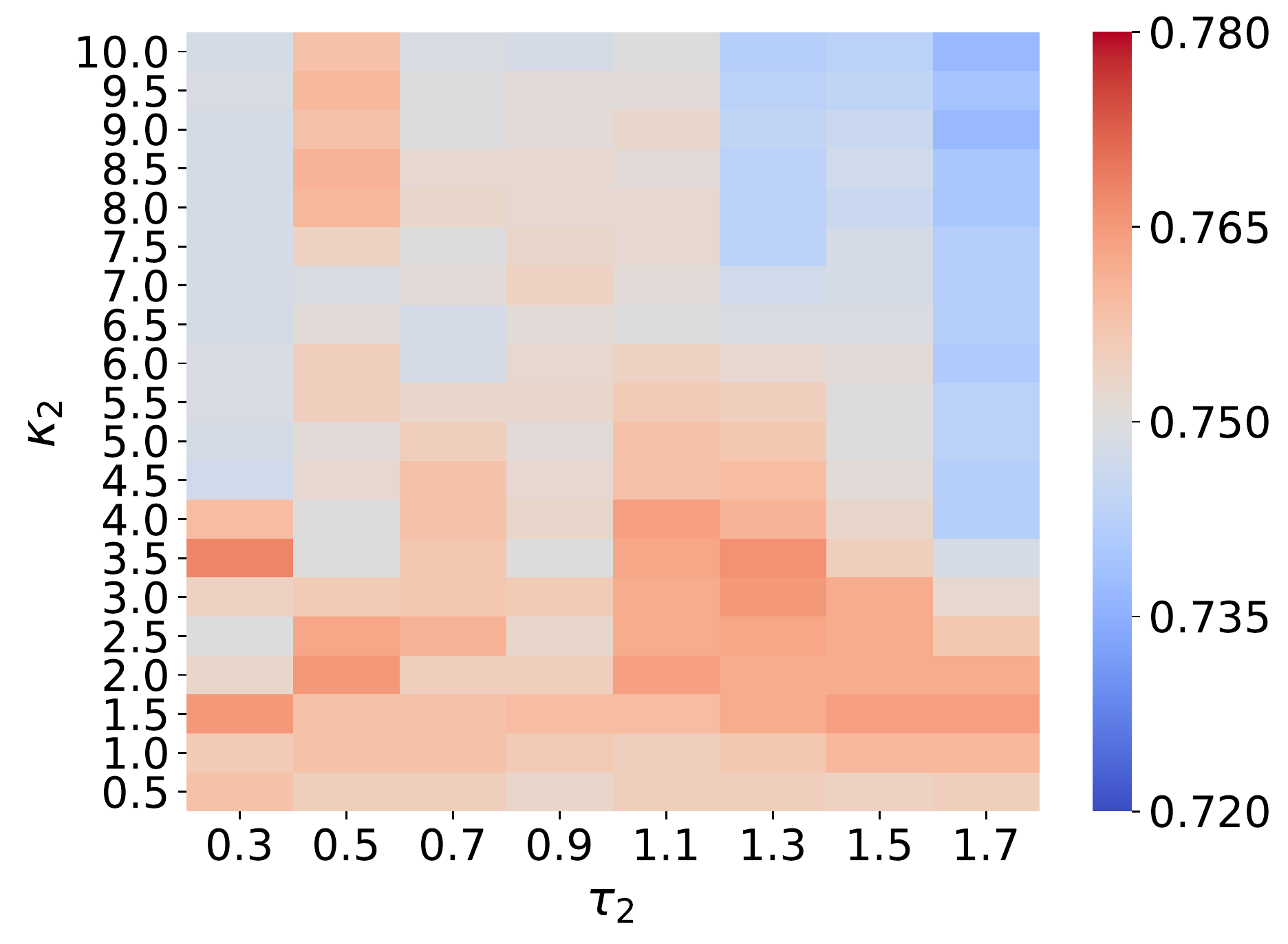}
		\caption{Minimum curvature} 
	\end{subfigure}
	\vspace{1em} 
	\begin{subfigure}{0.4\textwidth} 
		\includegraphics[width=\textwidth]{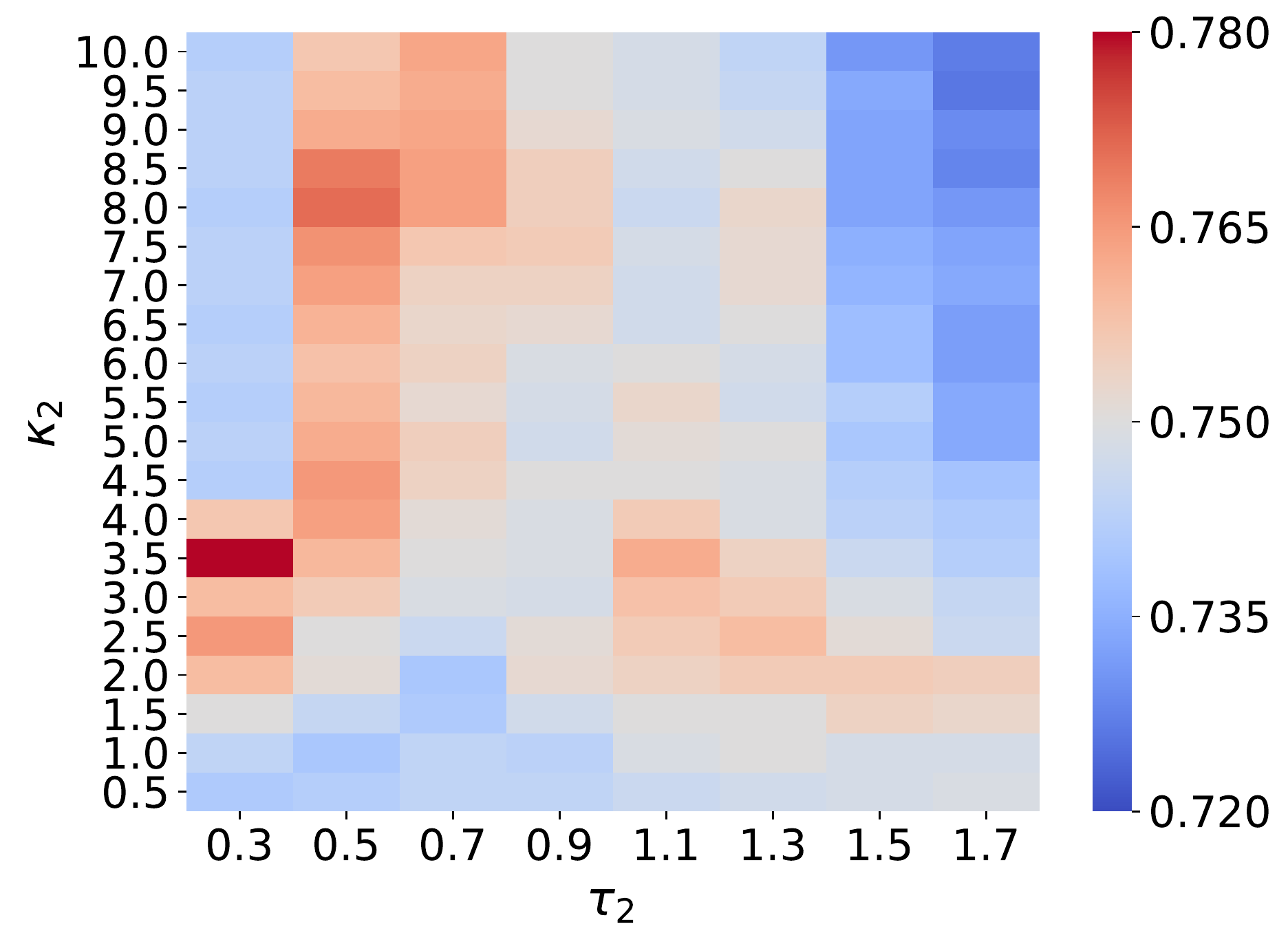}
		\caption{Maximum curvature} 
	\end{subfigure}
	\\
	\centering
	\begin{subfigure}{0.4\textwidth} 
		\includegraphics[width=\textwidth]{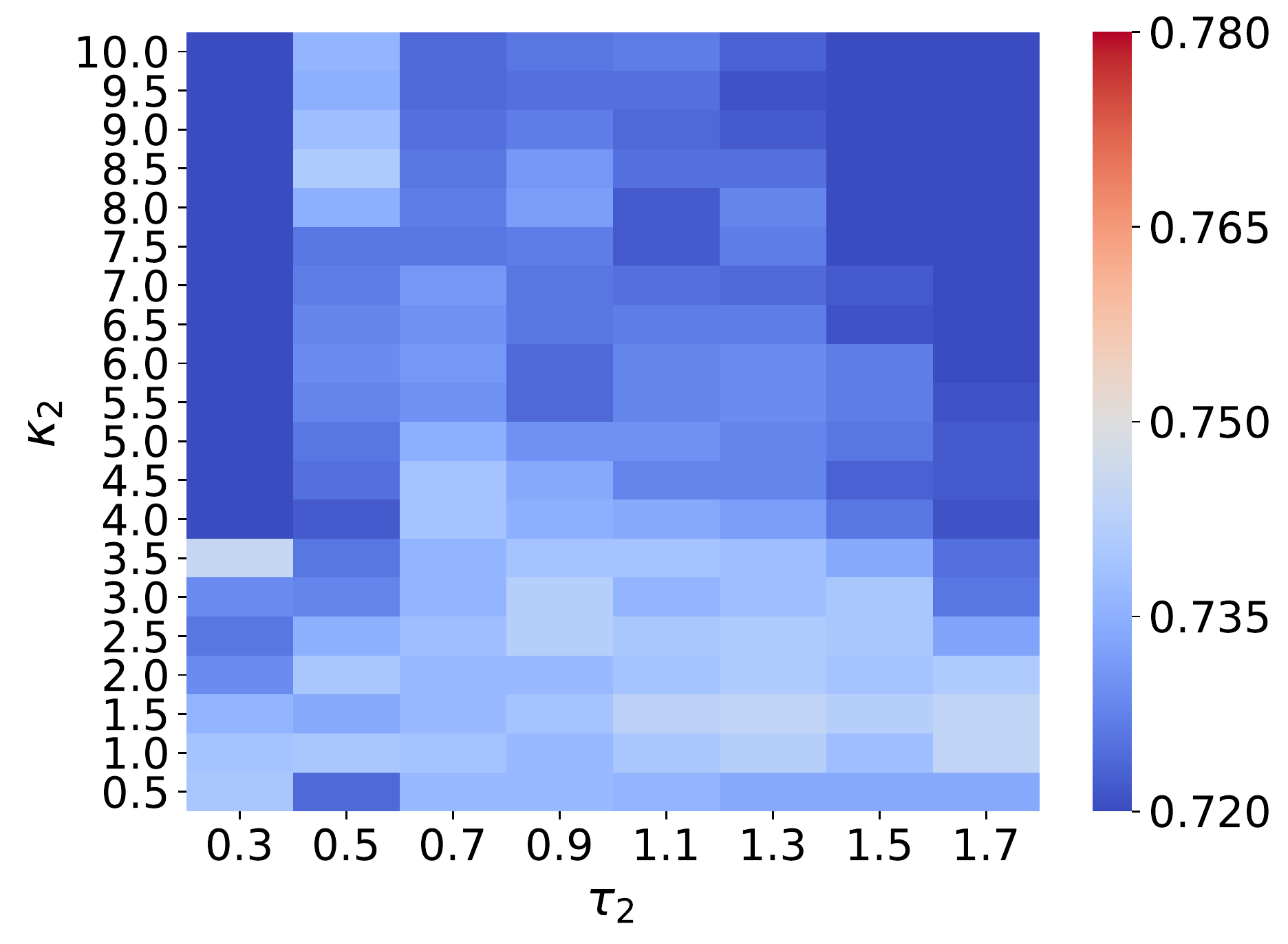}
		\caption{Gaussian curvature} 
	\end{subfigure}
	\vspace{1em} 
	\begin{subfigure}{0.4\textwidth} 
		\includegraphics[width=\textwidth]{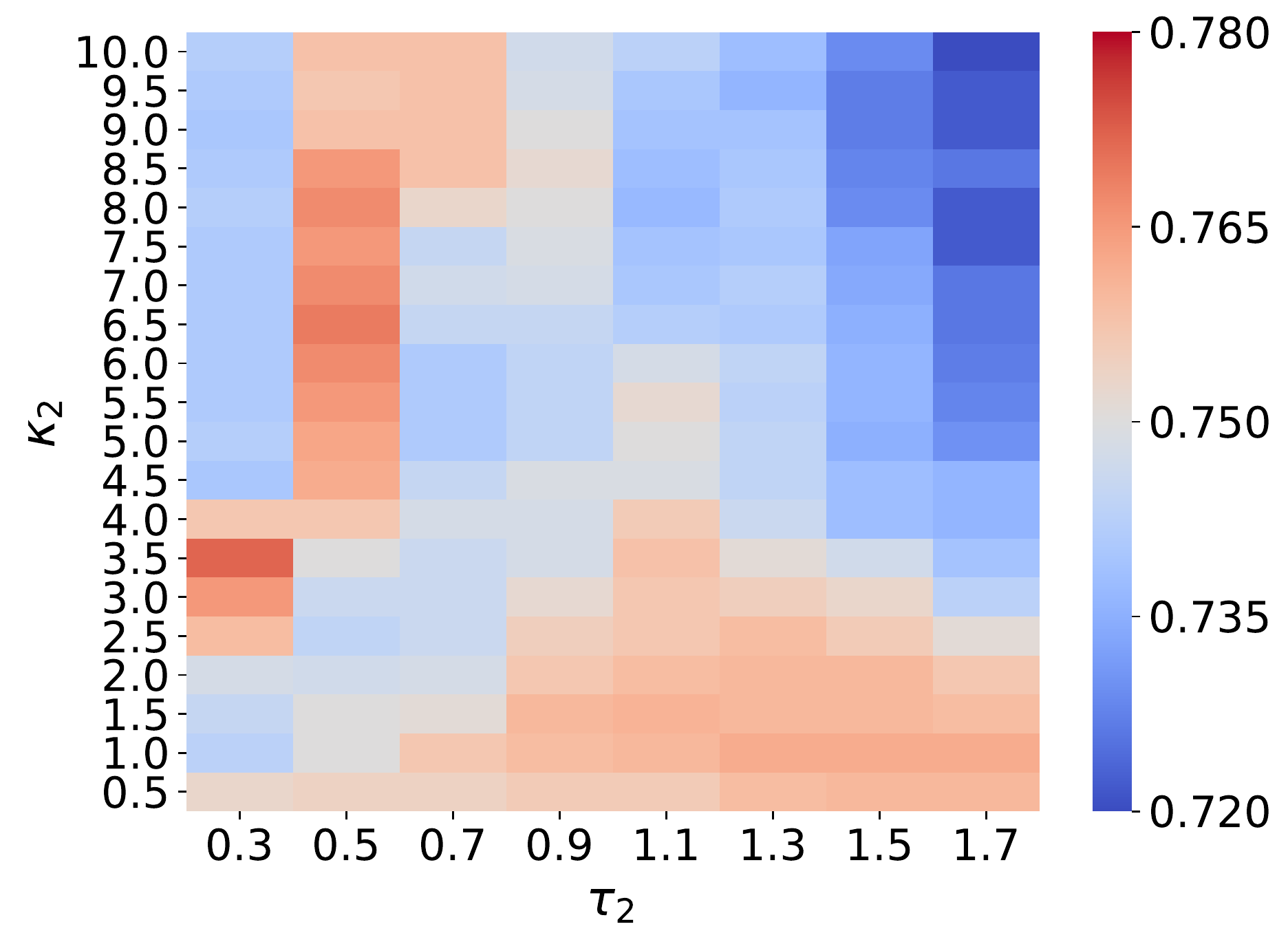}
		\caption{Mean curvature} 
	\end{subfigure}
	\caption{Median values of R-squared correlations $(R^2)$ from 4-fold cross validation performances of ${\rm EIC}^{CC}_{{\rm E}, \kappa_1,\tau_1; {\rm E}, \kappa_2,\tau_2}$ on the IGC$_{50}$ training set are plotted against different values of $\tau_2$ and $\kappa_2$. Two exponential kernels are utilized for features generation. While the parameters of the first kernel $(\tau_1, \kappa_1)$ are fixed and chosen from those reported in Fig. \ref{fig:tox-e-kfold}, the parameters of the second kernel  $(\tau_2, \kappa_2)$ are varied in the interested domains.
		The best performance for different kinds of curvatures is found as follows 
		(a) minimum curvature: $(\tau=0.7, \kappa=10)$, $(\tau_2=0.3, \kappa_2=3.5)$ with $R^2= 0.768$;
		(b) maximum curvature: $(\tau=0.3, \kappa=1.0)$, $(\tau_2=0.3, \kappa_2=3.5)$ with $R^2= 0.780$;
		(c) Gaussian curvature: $(\tau=0.7, \kappa=10)$, $(\tau_2=0.3, \kappa_2=3.5)$ with $R^2= 0.745$;
		(d) mean curvature $(\tau=0.3, \kappa=1.5)$, $(\tau_2=0.3, \kappa_2=3.5)$ with $R^2= 0.772$.}
	\label{fig:tox-ee-kfold}
\end{figure}

Figure \ref{fig:tox-ee-kfold} visualizes the 4-fold CV performances of ${\rm EIC}^{CC}_{{\rm E}, \kappa_1,\tau_1; {\rm E}, \kappa_2,\tau_2}$ on the IGC$_{50}$ training set against the different choices of $\kappa_2$ and $\tau_2$. Parameters for the first kernel $\kappa_1$ and $\tau_1$ are chosen from those reported in Fig. \ref{fig:tox-e-kfold}.

\begin{figure} [!ht]
	\centering
	\begin{subfigure}{0.4\textwidth} 
		\includegraphics[width=\textwidth]{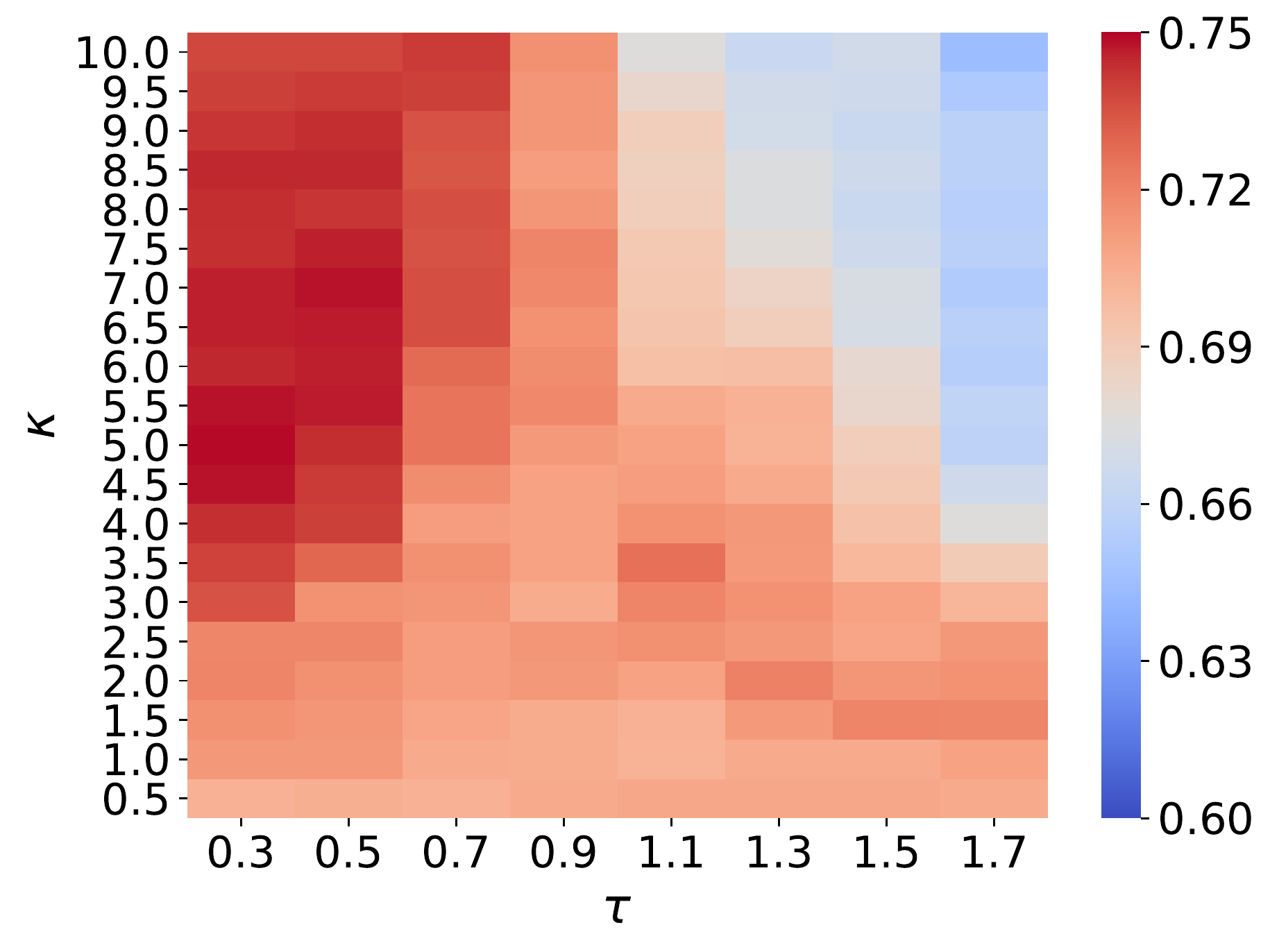}
		\caption{Minimum curvature} 
	\end{subfigure}
	\vspace{1em} 
	\begin{subfigure}{0.4\textwidth} 
		\includegraphics[width=\textwidth]{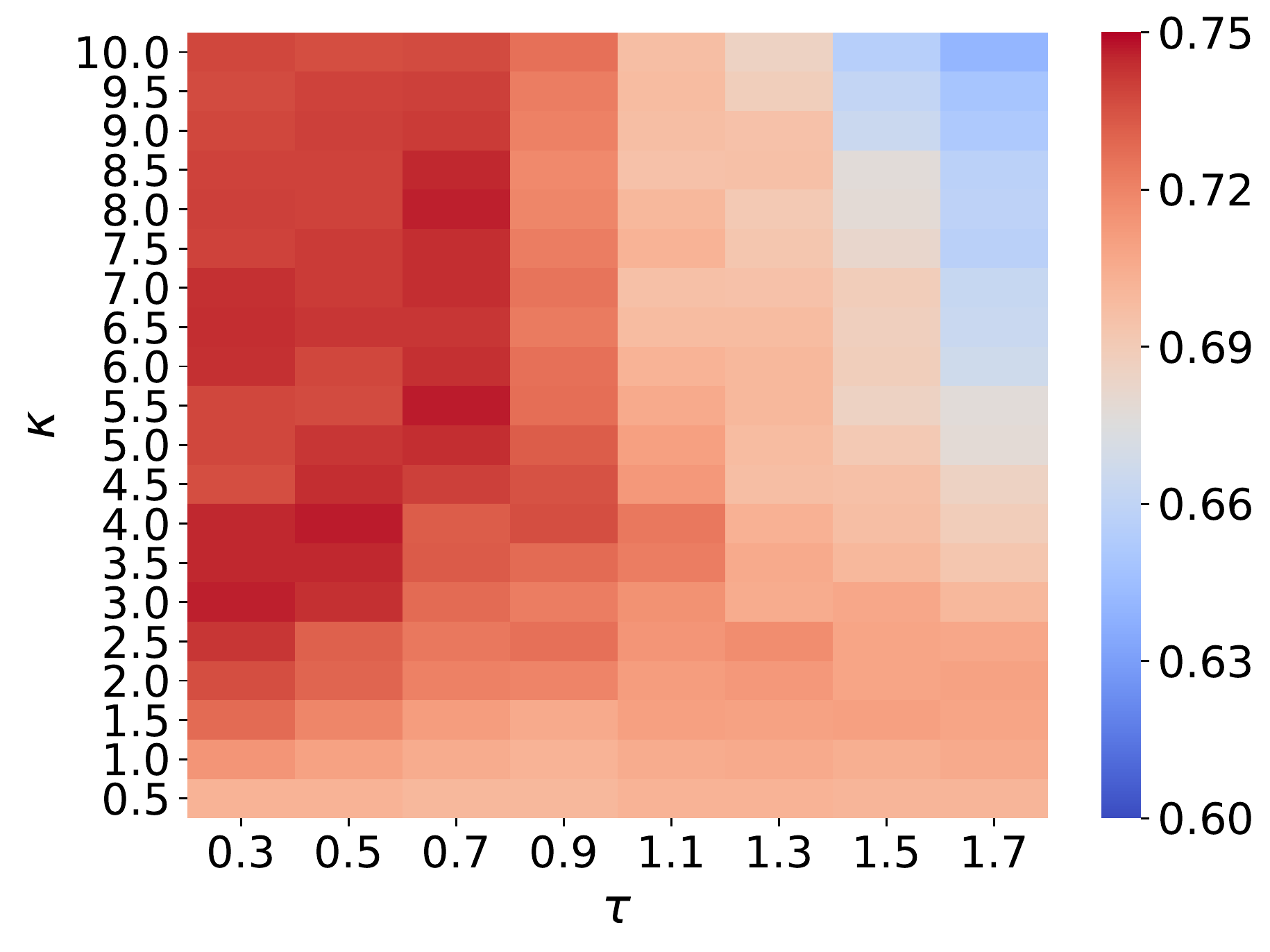}
		\caption{Maximum curvature} 
	\end{subfigure}
	\\
	\centering
	\begin{subfigure}{0.4\textwidth} 
		\includegraphics[width=\textwidth]{tox-exp-gauss-cv.eps}
		\caption{Gaussian curvature} 
	\end{subfigure}
	\vspace{1em} 
	\begin{subfigure}{0.4\textwidth} 
		\includegraphics[width=\textwidth]{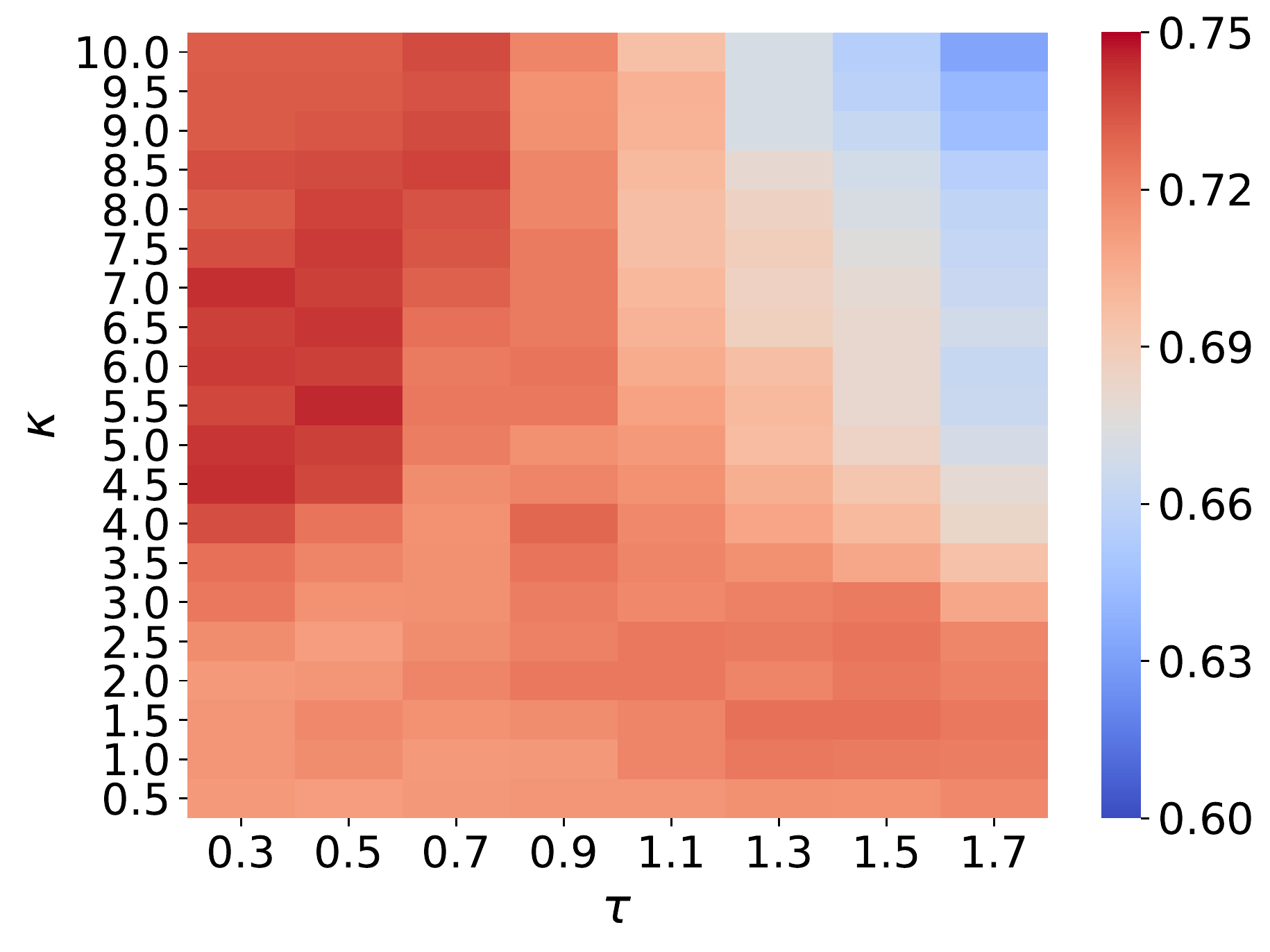}
		\caption{Mean curvature} 
	\end{subfigure}
	\caption{Median values of R-squared correlations $(R^2)$ from 4-fold cross validation performances of ${\rm EIC}^{C}_{{\rm L}, \kappa, \tau}$ on the IGC$_{50}$ training set are plotted against different values of $\tau$ and $\kappa$. Lorentz kernels are utilized for curvature features generation.
		The best performance for different kinds of curvatures is found as follows 
		(a) minimum curvature: $(\tau=0.3, \kappa=5.0)$ with $R^2= 0.749$;
		(b) maximum curvature: $(\tau=0.5, \kappa=4.0)$ with $R^2= 0.747$;
		(c) Gaussian curvature: $(\tau=0.3, \kappa=3.5)$ with $R^2= 0.724$;
		(d) mean curvature $(\tau=0.5, \kappa=5.5)$ with $R^2= 0.745$.}
	\label{fig:tox-l-kfold}
\end{figure}

 Figure \ref{fig:tox-l-kfold} illustrates the 4-fold cross-validation (CV) performance of ${\rm EIC}^{C}_{{\rm L}, \kappa,\tau}$ on the IGC$_{50}$ training set against the different choices of $\kappa$ and $\tau$.

\begin{figure} [!ht]
	\centering
	\begin{subfigure}{0.4\textwidth} 
		\includegraphics[width=\textwidth]{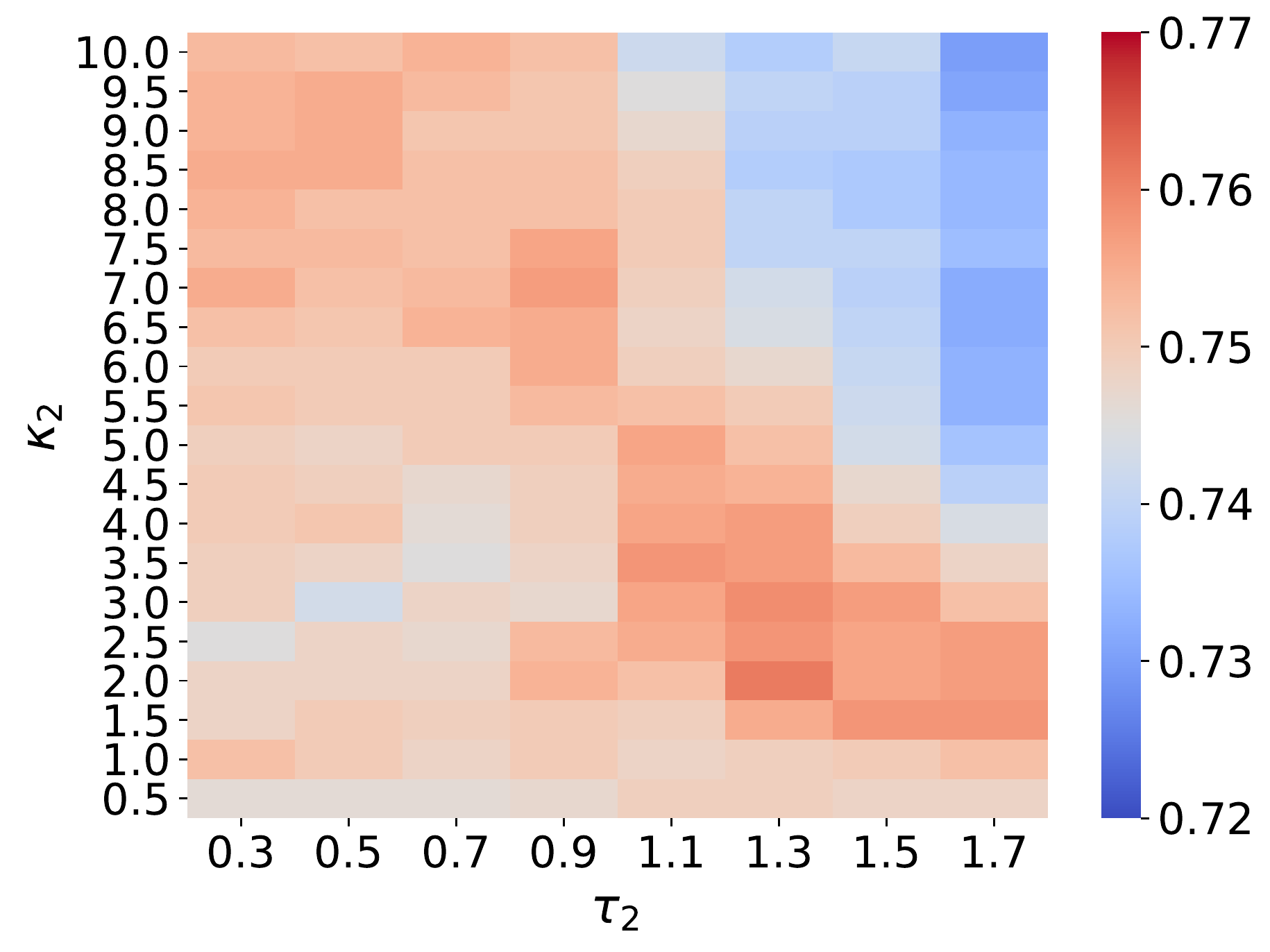}
		\caption{Minimum curvature} 
	\end{subfigure}
	\vspace{1em} 
	\begin{subfigure}{0.4\textwidth} 
		\includegraphics[width=\textwidth]{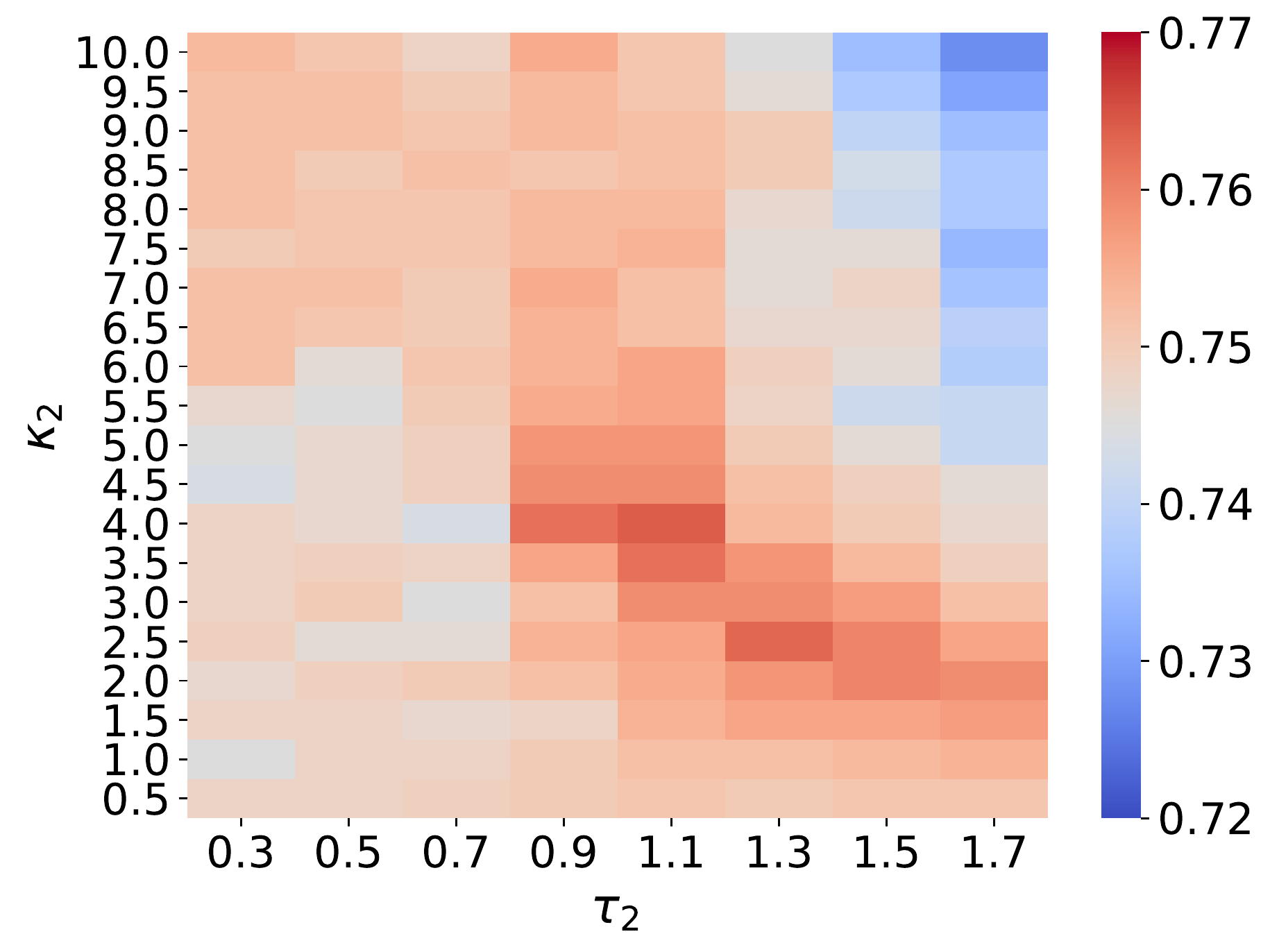}
		\caption{Maximum curvature} 
	\end{subfigure}
	\\
	\centering
	\begin{subfigure}{0.4\textwidth} 
		\includegraphics[width=\textwidth]{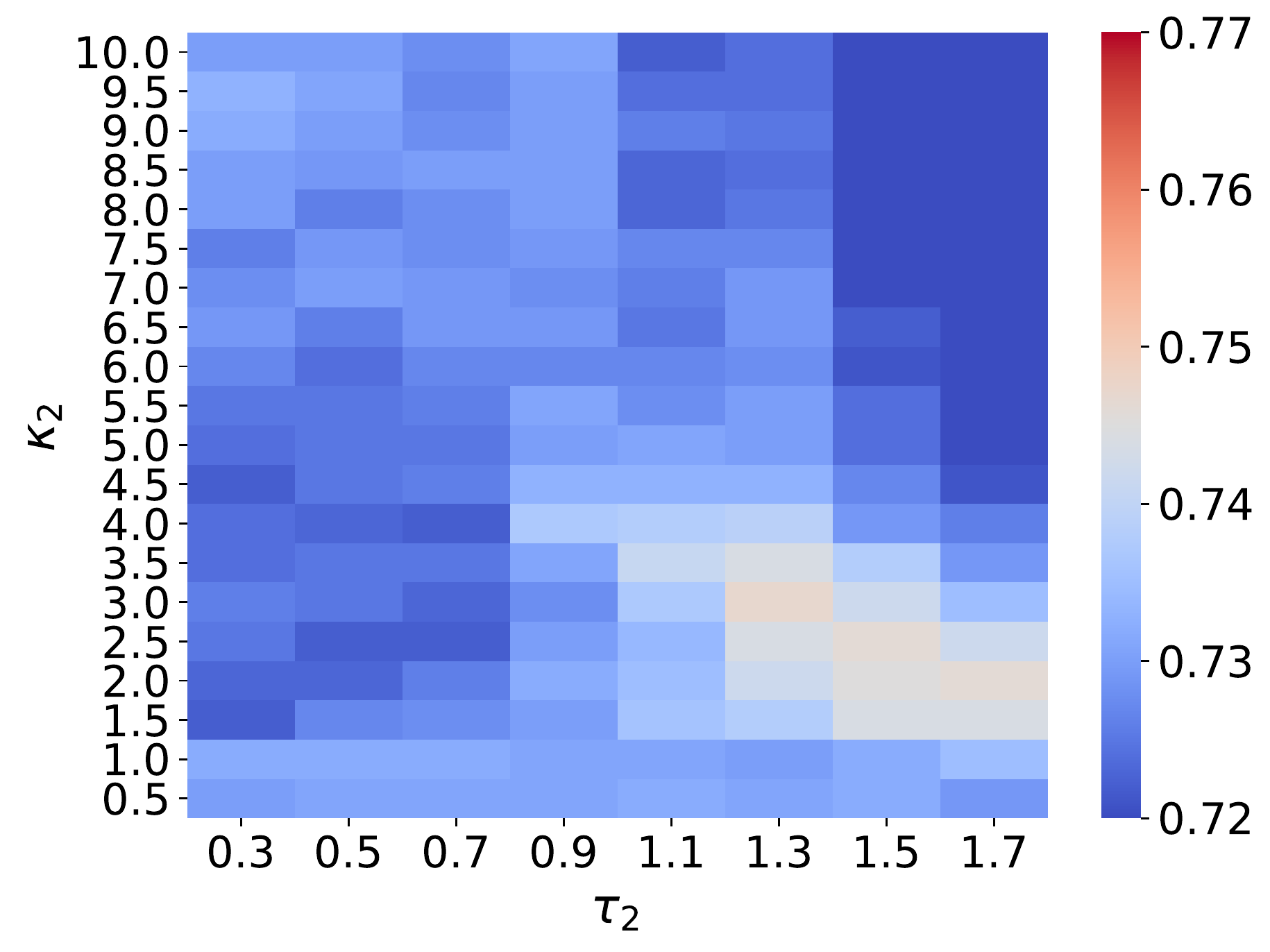}
		\caption{Gaussian curvature} 
	\end{subfigure}
	\vspace{1em} 
	\begin{subfigure}{0.4\textwidth} 
		\includegraphics[width=\textwidth]{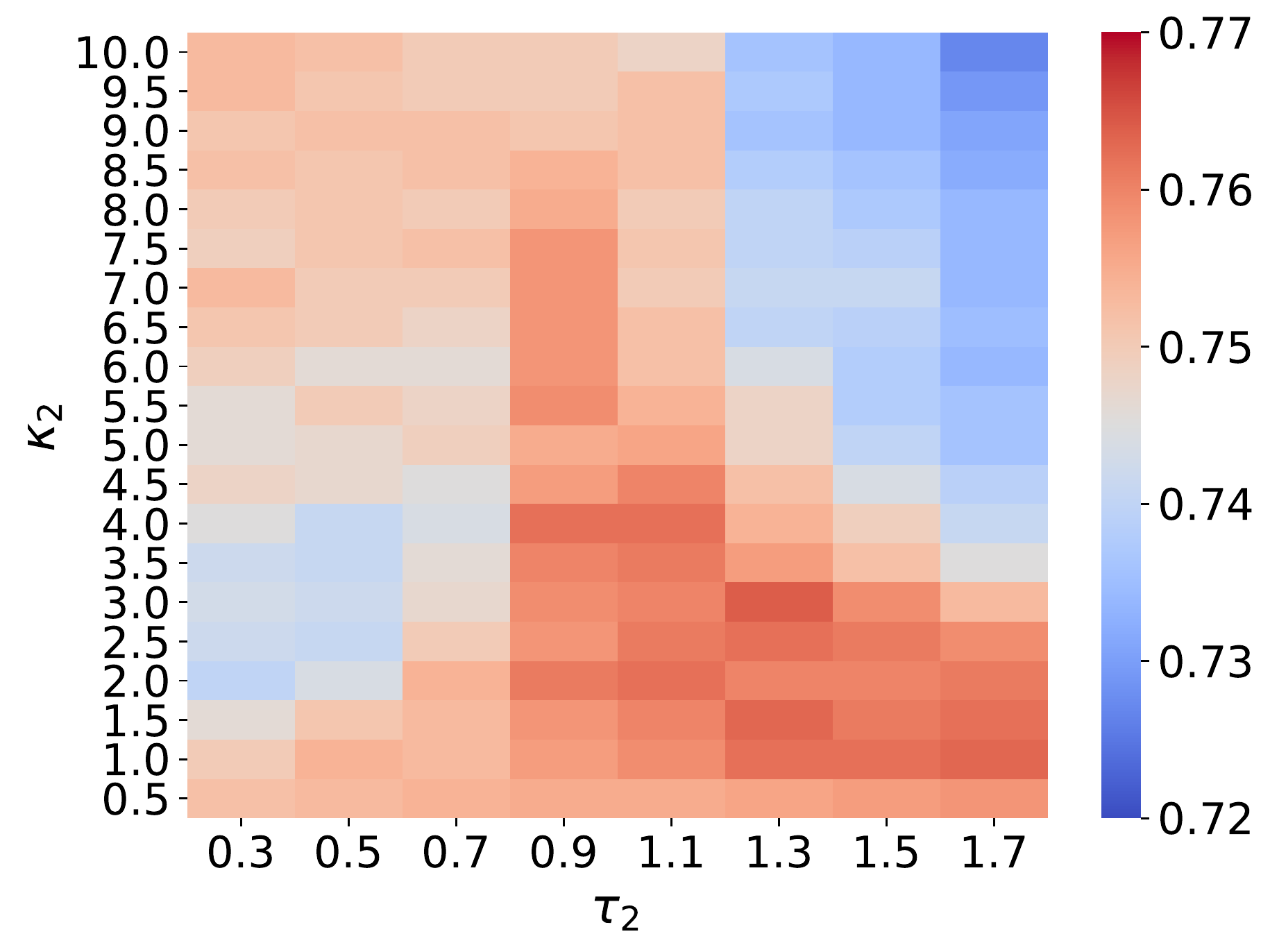}
		\caption{Mean curvature} 
	\end{subfigure}
	\caption{Median values of R-squared correlations $(R^2)$ from 4-fold cross validation performances of ${\rm EIC}^{CC}_{{\rm L}, \kappa_1,\tau_1; {\rm L}, \kappa_2,\tau_2}$ on the IGC$_{50}$ training set are plotted against different values of $\tau_2$ and $\kappa_2$. Two exponential kernels are utilized for features generation. While the parameters of the first kernel $(\tau_1, \kappa_1)$ are fixed and chosen from those reported in Fig. \ref{fig:tox-l-kfold}, the parameters of the second kernel  $(\tau_2, \kappa_2)$ are varied in the interested domains.
		The best performance for different kinds of curvatures is found as follows 
		(a) minimum curvature: $(\tau=0.3, \kappa=5.0)$, $(\tau_2=1.3, \kappa_2=2.0)$ with $R^2= 0.761$;
		(b) maximum curvature: $(\tau=0.5, \kappa=4.0)$, $(\tau_2=1.1, \kappa_2=4.0)$ with $R^2= 0.764$;
		(c) Gaussian curvature: $(\tau=0.3, \kappa=3.5)$, $(\tau_2=1.3, \kappa_2=3.0)$ with $R^2= 0.747$;
		(d) mean curvature $(\tau=0.5, \kappa=5.5)$, $(\tau_2=1.3, \kappa_2=3.0)$ with $R^2= 0.764$.}
	\label{fig:tox-ll-kfold}
\end{figure}

Figure \ref{fig:tox-ll-kfold} presents the 4-fold CV performances of ${\rm EIC}^{CC}_{{\rm L}, \kappa_1,\tau_1; {\rm L}, \kappa_2,\tau_2}$ on the IGC$_{50}$ training set against the different choices of $\kappa_2$ and $\tau_2$. Parameters for the first kernel $\kappa_1$ and $\tau_1$ are chosen from those reported in Fig. \ref{fig:tox-l-kfold}

\subsubsection{Parameter search using the test set}
Table \ref{tab:IGC50_results_testset} presents the prediction results of the IGC$_{50}$ test set with EIC models optimized by using the performance on the test set as the parameter selection target.
\begin{table*}[!ht]
	\centering
	\caption{Comparison of  prediction results for the Tetrahymena Pyriformis IGC$_{50}$ test set. Here, notation $^*$ indicates the EIC kernel parameters were optimized according to the test set performance.}
	\begin{tabular}{|c|c|c|c|c|c|c|}
		\hline
		Method & $R^2$ & $\frac{R^2-R_0^2}{R^2}$ & $k$ & RMSE & MAE & Coverage \\ \hline
		$^*$EIC$^{k_{\min}}_{E, 4, 0.5}$ & 0.766 & 0.002 & 0.999 & 0.476  & 0.351 & 1.000 \\
		EIC$^{k_{\min}k_{\min}}_{E, 4, 0.5; E, 8.5, 0.5}$ & 0.790 & 0.003 & 0.998 & 0.453  & 0.328 & 1.000  \\
		$^*$EIC$^{k_{\min}}_{L, 5, 0.3}$  & 0.759 & 0.002 & 1.000 & 0.484  & 0.339 & 1.000\\
		$^*$EIC$^{k_{\min}k_{\min}}_{L, 5, 0.3; L, 2, 1.7}$ & 0.782 & 0.004 & 0.999 & 0.462  & 0.320 & 1.000  \\
		$^*$Consensus$^{k_{\min}}$ &  0.799 & 0.006 & 1.000 &  0.445 &  0.315  & 1.000\\
		\hline
		$^*$EIC$^{k_{\max}}_{E, 4.5, 0.7}$   & 0.760 & 0.002 & 1.003 & 0.483 & 0.349 & 1.000 \\
		$^*$EIC$^{k_{\max}k_{\max}}_{E, 4.5, 0.7; E, 3.5, 0.3}$ & 0.780 & 0.002 & 0.999 & 0.463 & 0.339 & 1.000 \\
		$^*$EIC$^{k_{\max}}_{L, 7, 0.3}$ & 0.765 & 0.001 & 0.999 & 0.477 & 0.343 & 1.000 \\
		$^*$EIC$^{k_{\max}k_{\max}}_{L, 7, 0.3; L, 4, 1.3}$ & 0.783 & 0.003 & 0.999 & 0.460 & 0.324 & 1.000 \\
		Consensus$^{k_{\max}}$     & 0.792 & 0.004 & 1.000 & 0.452 & 0.323 & 1.000 \\
		\hline
		$^*$EIC$^{K}_{E, 2.5, 0.3}$  & 0.745 & 0.004 & 1.006 & 0.500 & 0.370 & 1.000\\
		$^*$EIC$^{KK}_{E, 2.5, 0.3; E, 1.5, 1.5}$ & 0.772 & 0.006 & 1.004 & 0.474 & 0.339 & 1.000 \\
		$^*$EIC$^{K}_{L, 2, 1.5}$  & 0.736 & 0.002 & 0.998 & 0.506 & 0.359 & 1.000 \\
		$^*$EIC$^{KK}_{L, 2, 1.5; L, 3, 0.3}$ & 0.768 & 0.005 & 1.000 & 0.477 & 0.343 & 1.000 \\
		$^*$Consensus$^K$   & 0.777 & 0.007 & 1.003 & 0.469 & 0.334 & 1.000  \\
		\hline
		$^*$EIC$^{H}_{E, 4, 0.5}$   & 0.765 & 0.003 & 1.001 & 0.478 & 0.349 & 1.000\\
		$^*$EIC$^{HH}_{E, 4, 0.5; E, 9.5, 0.5}$  & 0.783 & 0.002 & 0.999 & 0.460 & 0.336 & 1.000  \\
		$^*$EIC$^{H}_{L, 5.5, 0.3}$   & 0.759 & 0.002 & 0.998 & 0.484 & 0.343 & 1.000 \\
		$^*$EIC$^{HH}_{L, 5.5, 0.3; L, 3.0, 1.1}$   & 0.783 & 0.004 & 1.001 & 0.461 & 0.319 & 1.000  \\
		$^*$Consensus$^H$      & 0.797 & 0.005 & 1.001 & 0.447 & 0.319 & 1.000 \\ \hline 
		
	\end{tabular}
	\label{tab:IGC50_results_testset}
\end{table*}

\subsection{Solvation energy prediction}
\subsubsection{Parameter search using the training set cross-validation}
\begin{figure}[!ht]
	\centering
	\begin{subfigure}{0.4\textwidth} 
		\includegraphics[width=\textwidth]{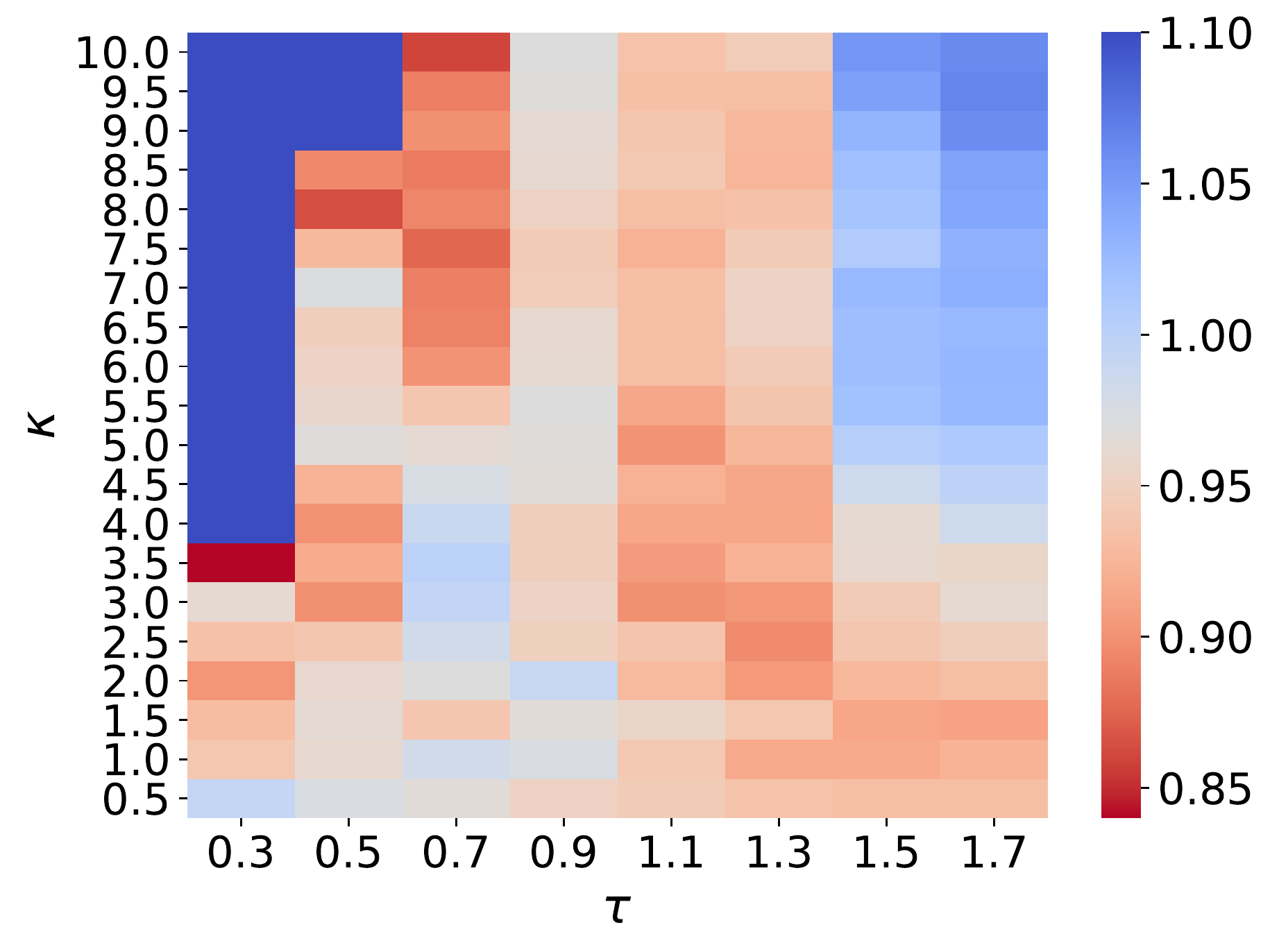}
		\caption{Mean curvature (exponential kernel)} 
	\end{subfigure}
	\vspace{1em} 
	\begin{subfigure}{0.4\textwidth} 
		\includegraphics[width=\textwidth]{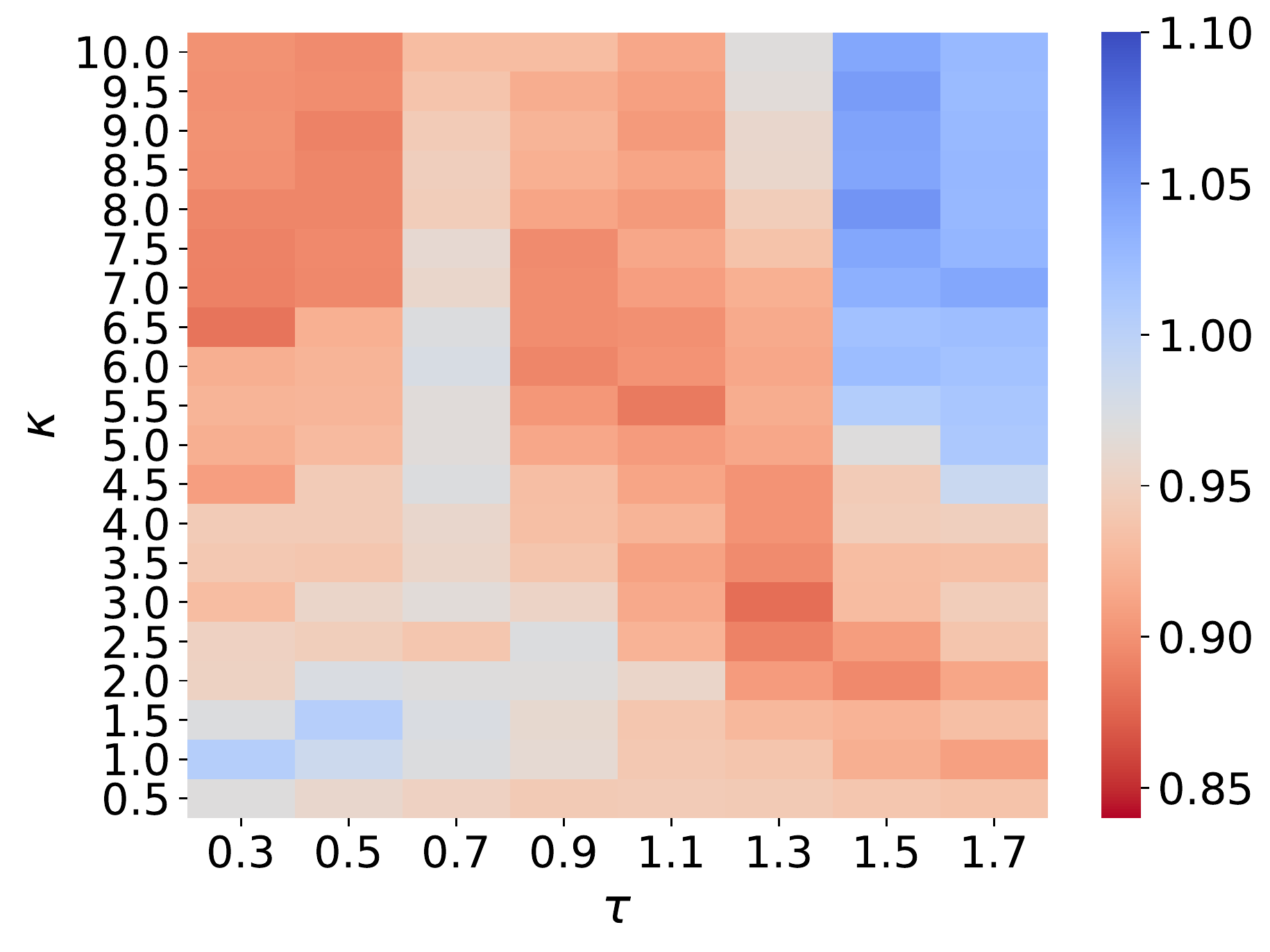}
		\caption{Mean curvature (Lorentz kernel)} 
	\end{subfigure}
	\caption{Mean absolute error (MAE) from 3-fold cross-validation of  ${\rm EIC}^{H}_{{\rm \alpha}, \kappa,\tau}$ on the solvation training set are plotted against different values of $\tau$ and $\kappa$. The element interactive  mean curvatures are utilized for all calculations.
		The best parameters and median values of MAE for  each model are found to be
		(a) exponential-kernel model: $(\tau=0.3, \kappa=3.5, {\rm MAE}=0.840)$;
		(b) Lorentz-kernel model: $(\tau=1.3, \kappa=3.0, {\rm MAE}=0.880)$
	}
	\label{fig:sol-el-kfold}
\end{figure}

We here use 3-fold cross-validation on the training dataset to select the best parameters. Fig. \ref{fig:sol-el-kfold} illustrates the 3-fold CV performance of ${\rm EIC}^{H}_{{\rm \alpha}, \kappa,\tau}$ on the solvation training set against the different choices of $\kappa$ and $\tau$.

\begin{figure}[!ht]
	\centering
	\begin{subfigure}{0.4\textwidth} 
		\includegraphics[width=\textwidth]{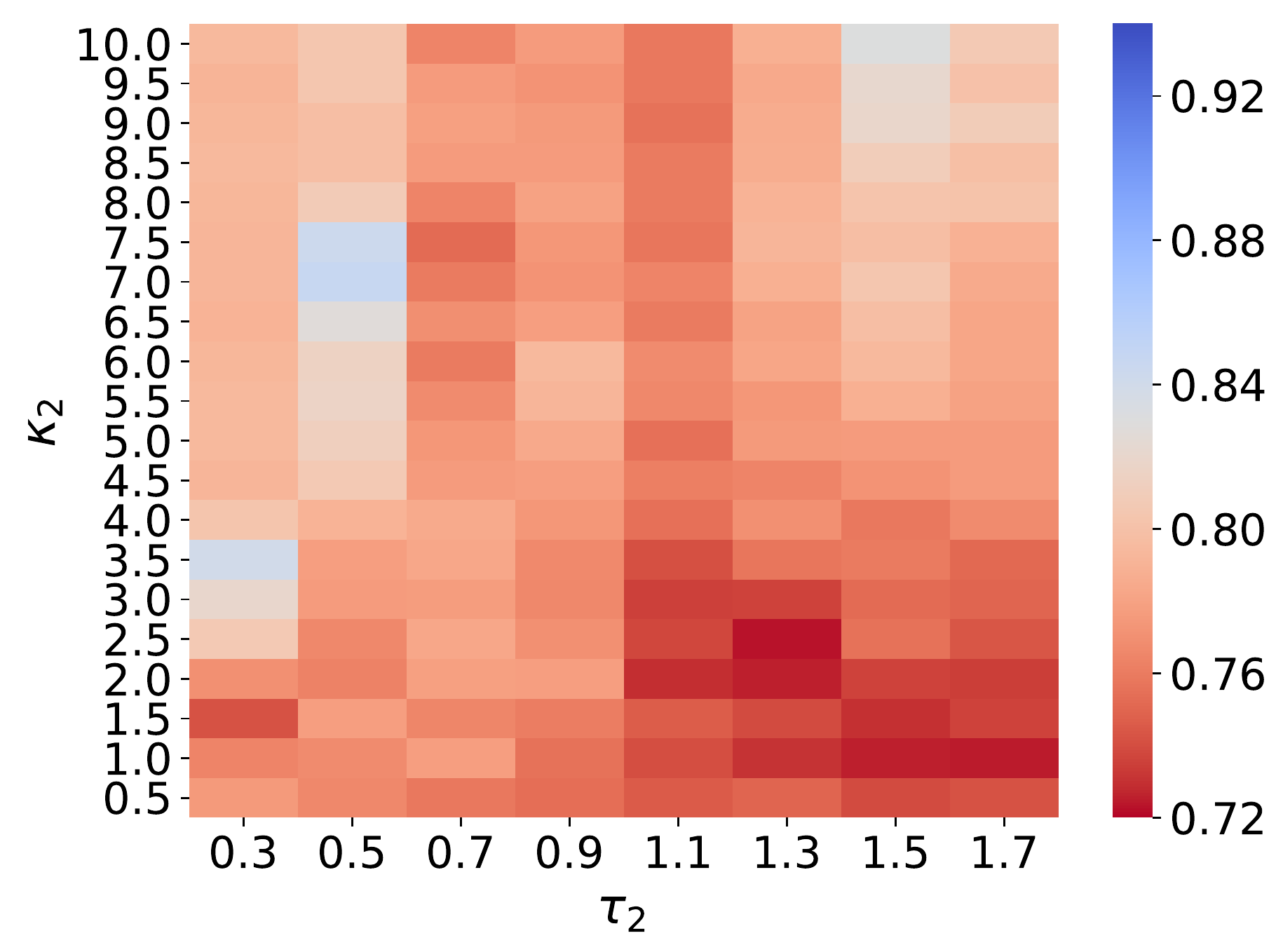}
		\caption{Mean curvature (exponential kernels)} 
	\end{subfigure}
	\vspace{1em} 
	\begin{subfigure}{0.4\textwidth} 
		\includegraphics[width=\textwidth]{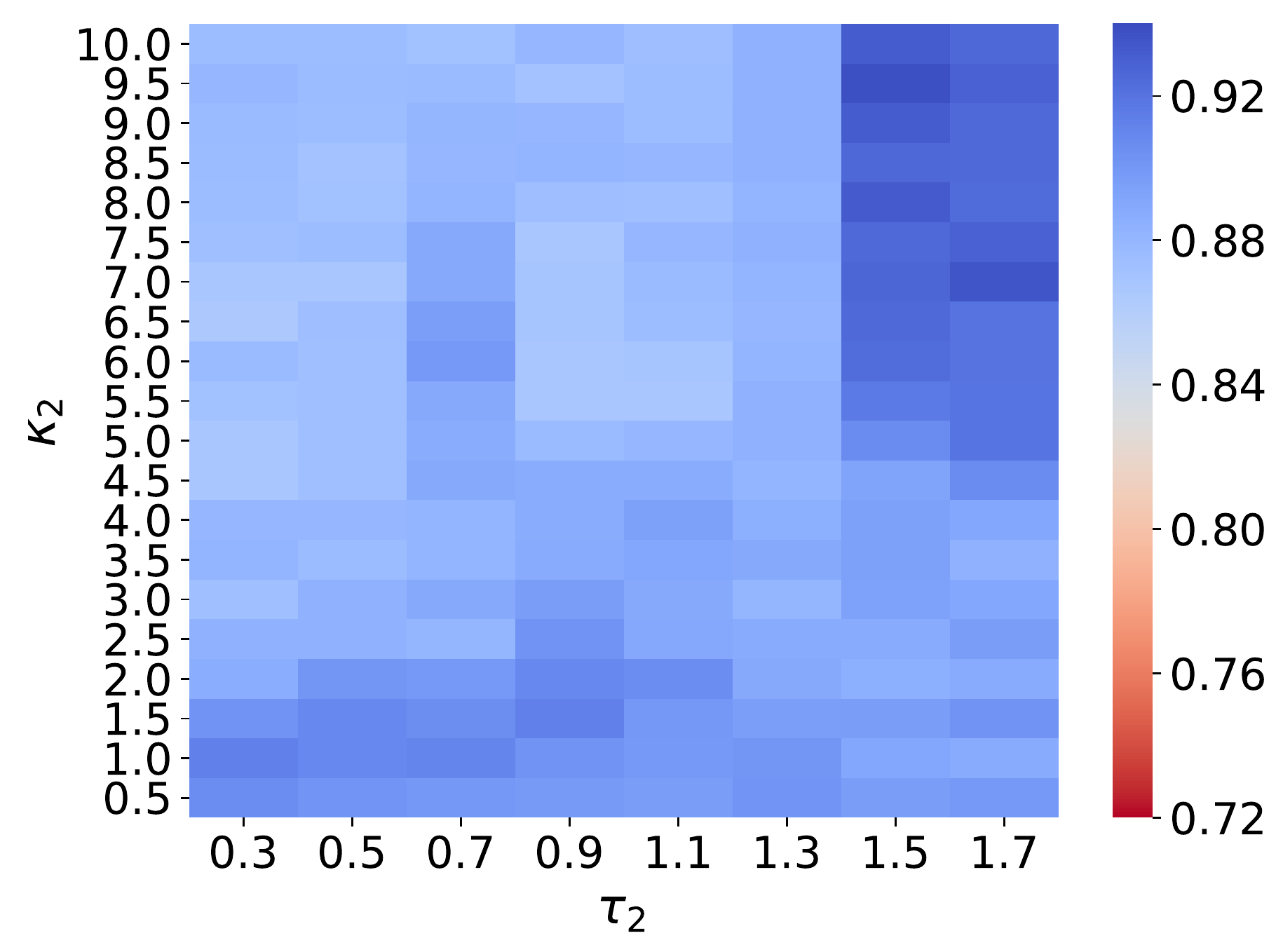}
		\caption{Mean curvature (Lorentz kernels)} 
	\end{subfigure}
	\caption{Mean absolute errors (MAEs) from 3-fold cross-validation  of ${\rm EIC}^{HH}_{{\alpha}, \kappa_1,\tau_1; {\alpha}, \kappa_2,\tau_2}$ on the solvation training set are plotted against different values of $\tau_2$ and $\kappa_2$. The element interactive  mean curvatures are utilized for all calculations. While the parameters of the first kernel $(\tau_1, \kappa_1)$ are fixed, the parameters of the second kernel  $(\tau_2, \kappa_2)$ are varied in the interested domains. The best parameters and median values of MAE for each model are found to be 
		(a) exponential-kernel model: $(\tau_1=0.3, \kappa_1=3.5, \tau_2=1.3, \kappa_2=2.5, {\rm MAE}=0.723)$;
		(b) Lorentz-kernel model: $(\tau_1=1.3, \kappa_1=3.0, \tau_2=0.3, \kappa_2=6.5, {\rm MAE}=0.866)$
	}
	\label{fig:sol-eell-kfold}
\end{figure}

Figure \ref{fig:sol-eell-kfold} presents the 3-fold cv performances of ${\rm EIC}^{HH}_{{\alpha}, \kappa_1,\tau_1; {\alpha}, \kappa_2,\tau_2}$ on the solvation training set against the different choices of $\kappa_2$ and $\tau_2$. Parameters for the first kernel $\kappa_1$ and $\tau_1$ are chosen from the report in Fig. \ref{fig:sol-el-kfold}

\subsubsection{Parameter search using the test set}

Table \ref{solvation_results_testset} presents the prediction results of the solvation test set with EIC models optimized by using the performance on test set as the parameter selection target.
\begin{table*}[!ht]
	\centering
	\caption{Comparison of  prediction results for the solvation dataset collected by Wang {\it et al} \cite{wangjm:2001}. Here, notation $^*$ indicates the EIC kernel parameters were optimized according to the test set performance.}
	\begin{tabular}{|c|c|c|c|}
		\hline
		Method & MAE (kcal/mol) & RMSE (kcal/mol) & $R^2$\\ \hline
		$^*$EIC$^H_{{\rm E},3.5, 0.3}$   & 0.575 & 0.921 & 0.904 \\
		$^*$EIC$^{HH}_{{\rm E},3.5, 0.3;{\rm E},4.0,1.3}$   & 0.518 & 0.812 & 0.929 \\
		$^*$EIC$^H_{{\rm L}, 5, 0.3}$    & 0.579 & 0.862 & 0.917 \\
		$^*$EIC$^{HH}_{{\rm L}, 5, 0.3; {\rm L},0.5,0.9}$   & 0.559 & 0.842 & 0.922 \\
		$^*$Consensus$^H$      & 0.524 & { 0.798} & { 0.931} \\
		\hline 
	\end{tabular}
	\label{solvation_results_testset}
\end{table*}

\subsection{Protein-ligand binding affinity prediction}
\subsubsection{Parameter search using the training set cross-validation}
\begin{figure}[!ht]
	\centering
	\begin{subfigure}{0.4\textwidth} 
		\includegraphics[width=\textwidth]{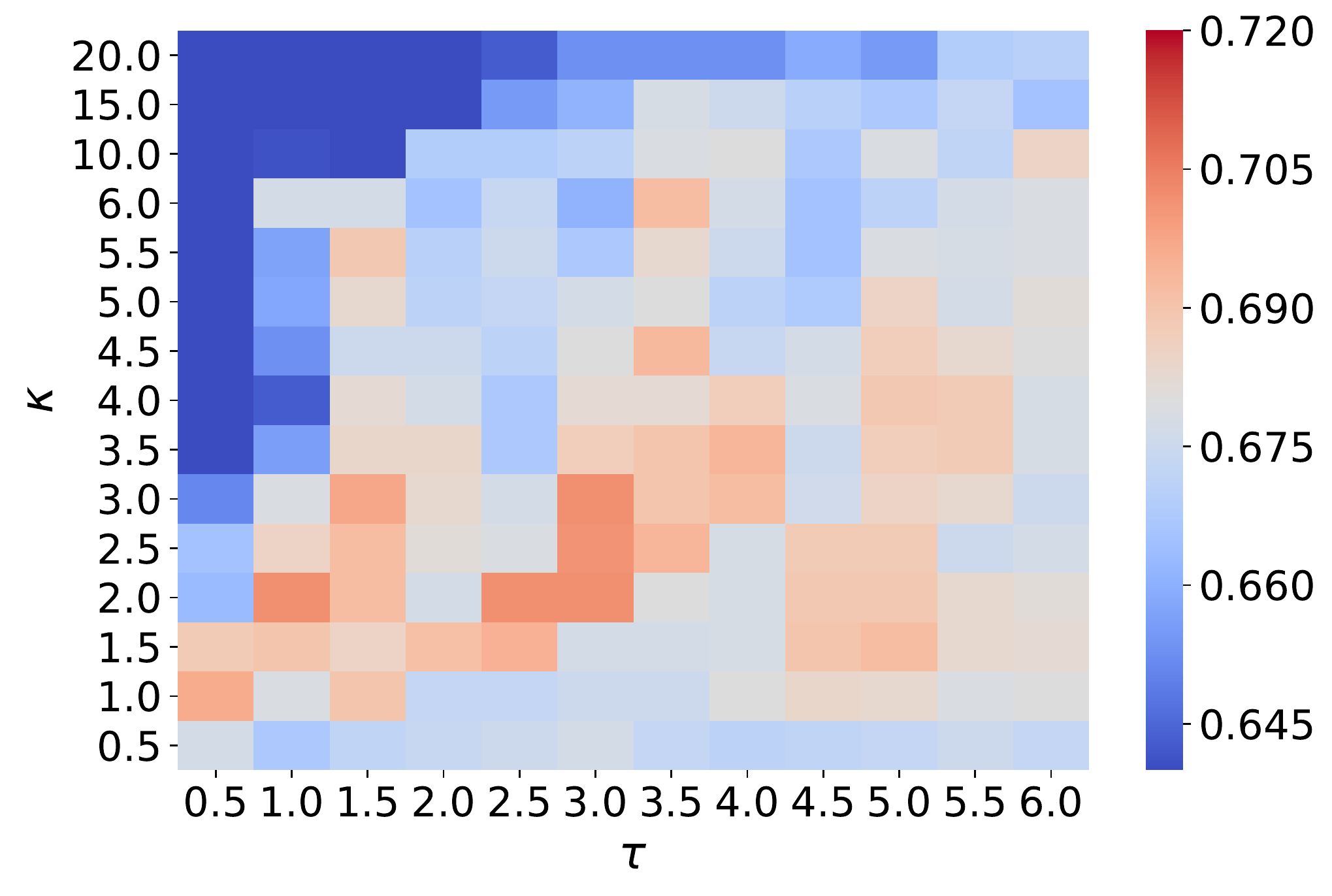}
		\caption{Mean curvature (exponential kernel)} 
	\end{subfigure}
	\vspace{1em} 
	\begin{subfigure}{0.4\textwidth} 
		\includegraphics[width=\textwidth]{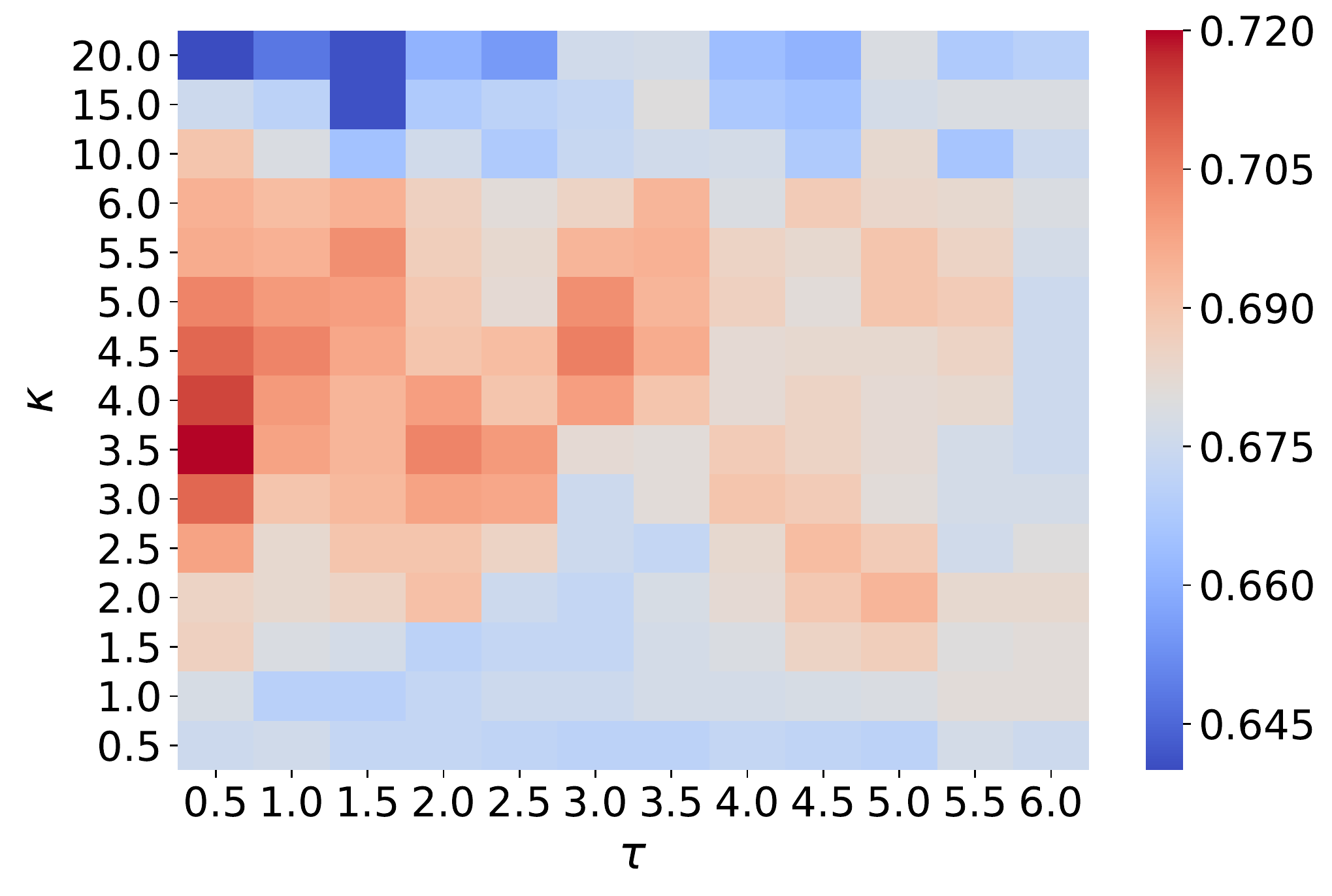}
		\caption{Mean curvature (Lorentz kernel)} 
	\end{subfigure}
	\caption{Pearson correlation coefficients $(R_p)$ from 5-fold cross-validation of ${\rm EIC}^{H}_{{\alpha}, \kappa,\tau}$ on the PDBbind v2007 refined set (excluding the core set) are plotted against different values of $\tau$ and $\kappa$. The element interactive  mean curvatures are utilized for all calculations.
		The best parameters and median values of $R_p$ for  ${\rm EIC}^{H}_{{\alpha}, \kappa,\tau}$ are respectively found to be 
		(a) exponential-kernel model: $(\tau=1.0, \kappa=2.0, R_p=0.702)$;
		(b) Lorentz-kernel model: $(\tau=0.5, \kappa=3.5, R_p=0.720)$
	}
	\label{fig:v2007-el-kfold}
\end{figure}

Figure \ref{fig:v2007-el-kfold} illustrates the 5-fold CV performance of ${\rm EIC}^{H}_{{\alpha}, \kappa,\tau}$ on the training set of the PDBbind v2007 benchmark against the different choices of $\kappa$ and $\tau$.

\begin{figure}
	\centering
	\begin{subfigure}{0.4\textwidth} 
		\includegraphics[width=\textwidth]{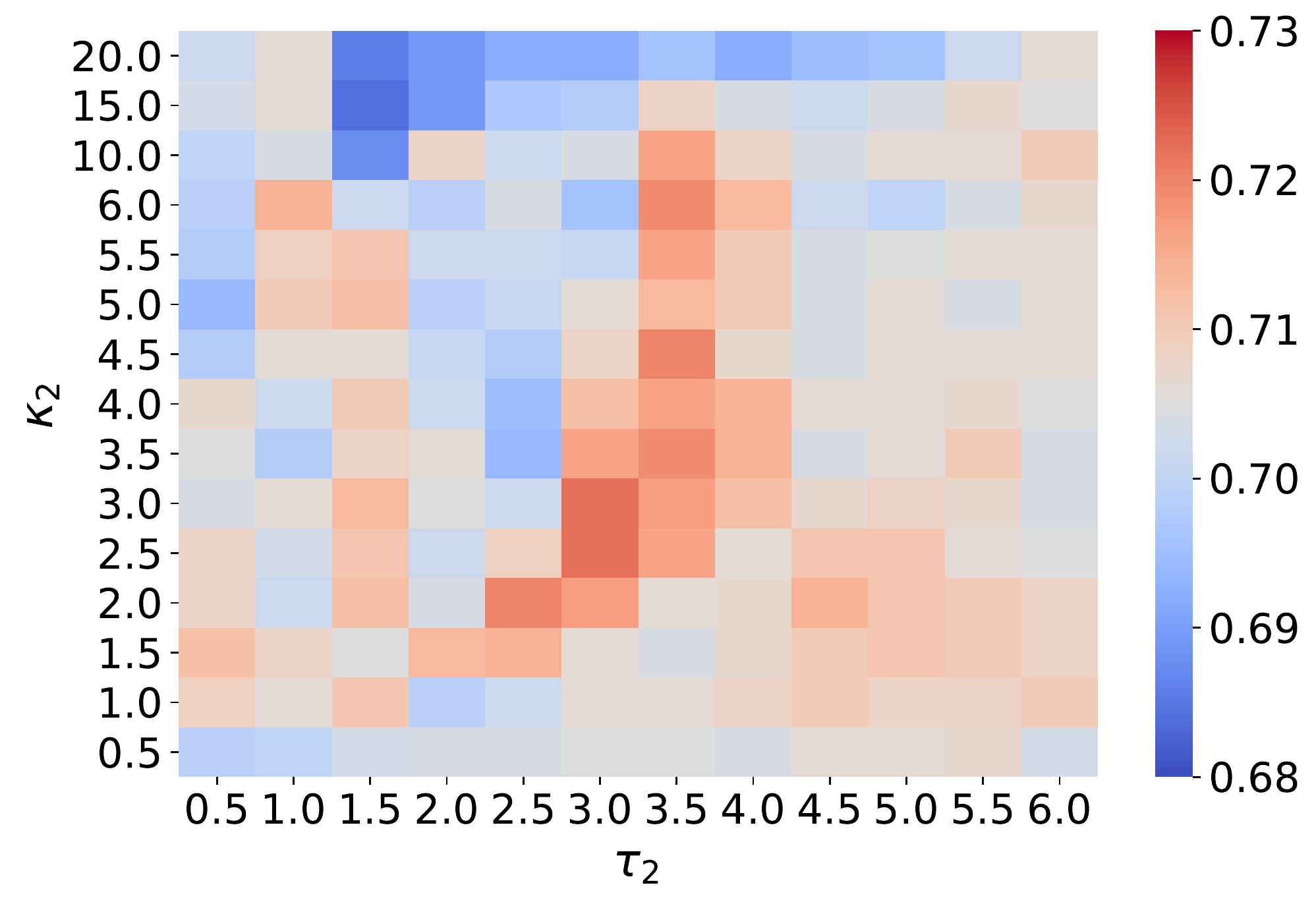}
		\caption{Mean curvature (exponential kernel)} 
	\end{subfigure}
	\vspace{1em} 
	\begin{subfigure}{0.4\textwidth} 
		\includegraphics[width=\textwidth]{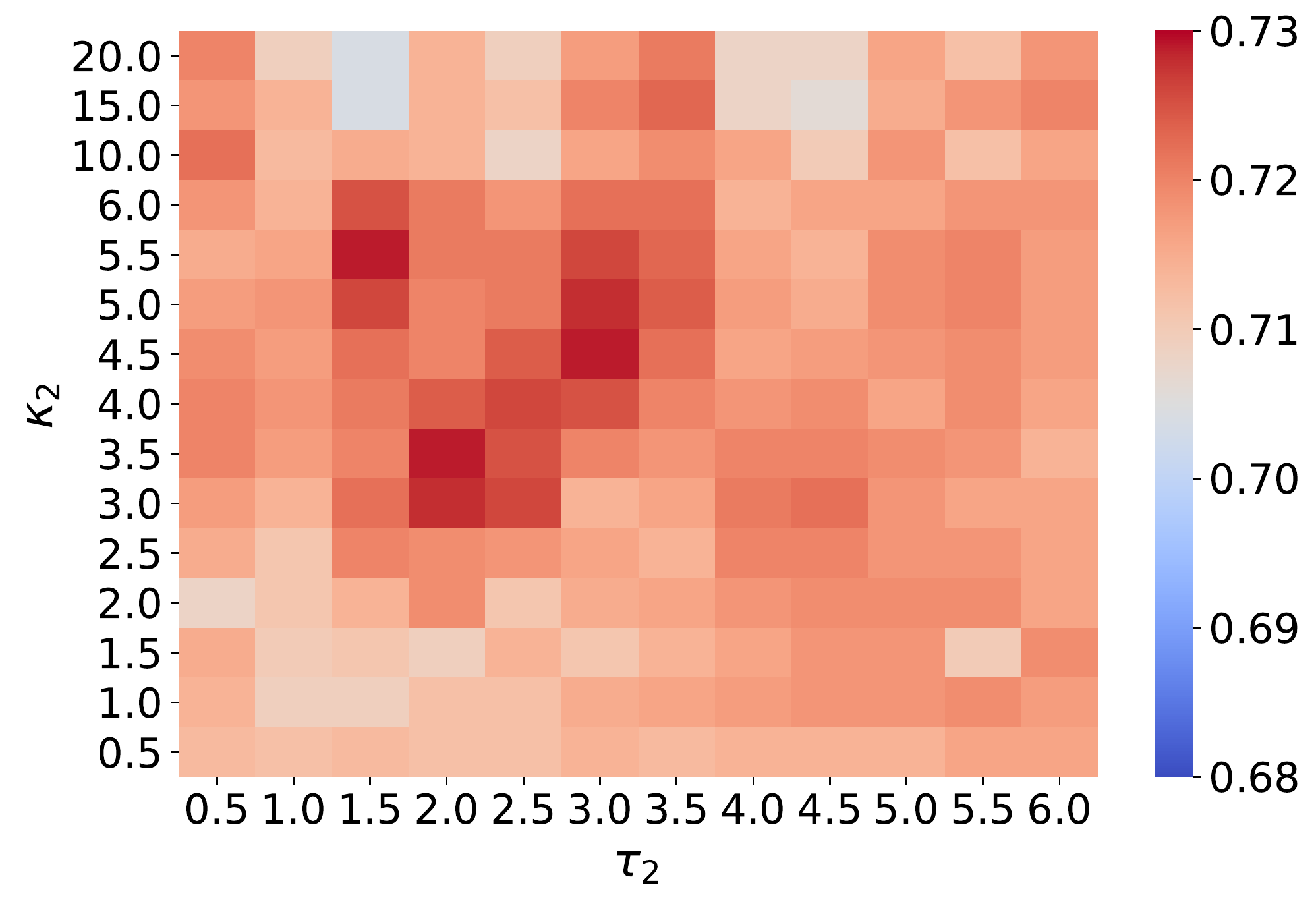}
		\caption{Mean curvature (Lorentz kernel)} 
	\end{subfigure}
	\caption{Pearson correlation coefficients  $(R_p)$ from 5-fold cross-validation of ${\rm EIC}^{HH}_{{\alpha}, \kappa_1,\tau_1; {\alpha}, \kappa_2,\tau_2}$ on the PDBbind v2007 refined set (excluding the core set) are plotted against different values of $\tau_2$ and $\kappa_2$. The element interactive  mean curvatures are utilized for all calculations. While the parameters of the first kernel $(\tau_1, \kappa_1)$ are fixed, the parameters of the second kernel  $(\tau_2, \kappa_2)$ are varied in the interested domains. The best parameters and median values of $R_p$ for ${\rm EIC}^{HH}_{{\alpha}, \kappa_1,\tau_1; {\alpha}, \kappa_2,\tau_2}$ are respectively found to be 
		(a) exponential-kernel model: $(\tau_1=1.0, \kappa_1=2.0, \tau_2=3.0, \kappa_2=3.0 R_p=0.722)$;
		(b) Lorentz-kernel model: $(\tau_1=0.5, \kappa_1=3.5, \tau_2=2.0, \kappa_2=3.5 R_p=0.729)$
	}
	\label{fig:v2007-eell-kfold}
\end{figure}

\begin{figure}[!ht]
	\centering
	\begin{subfigure}{0.4\textwidth} 
		\includegraphics[width=\textwidth]{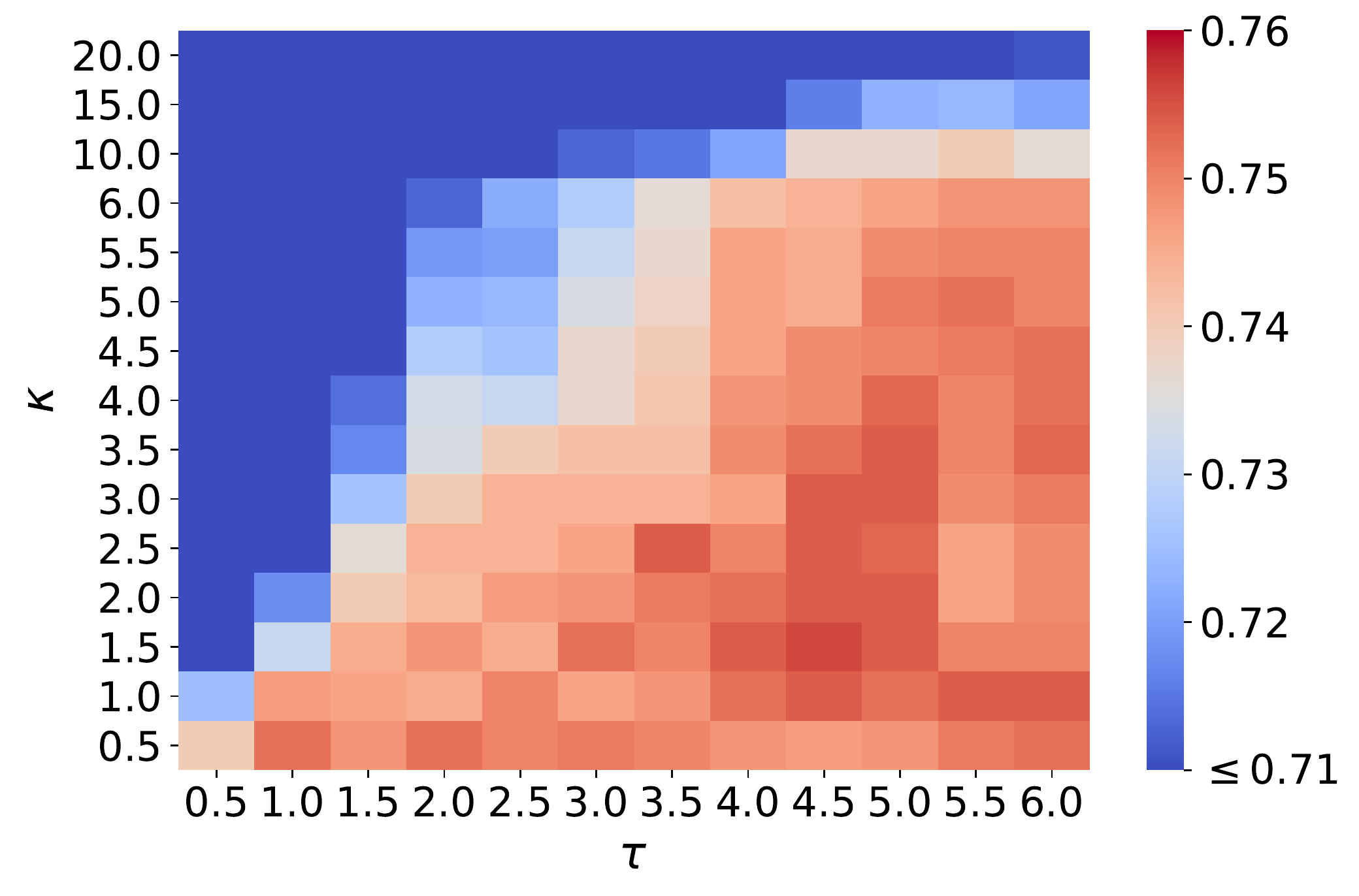}
		\caption{Mean curvature (exponential kernel)} 
	\end{subfigure}
	\vspace{1em} 
	\begin{subfigure}{0.4\textwidth} 
		\includegraphics[width=\textwidth]{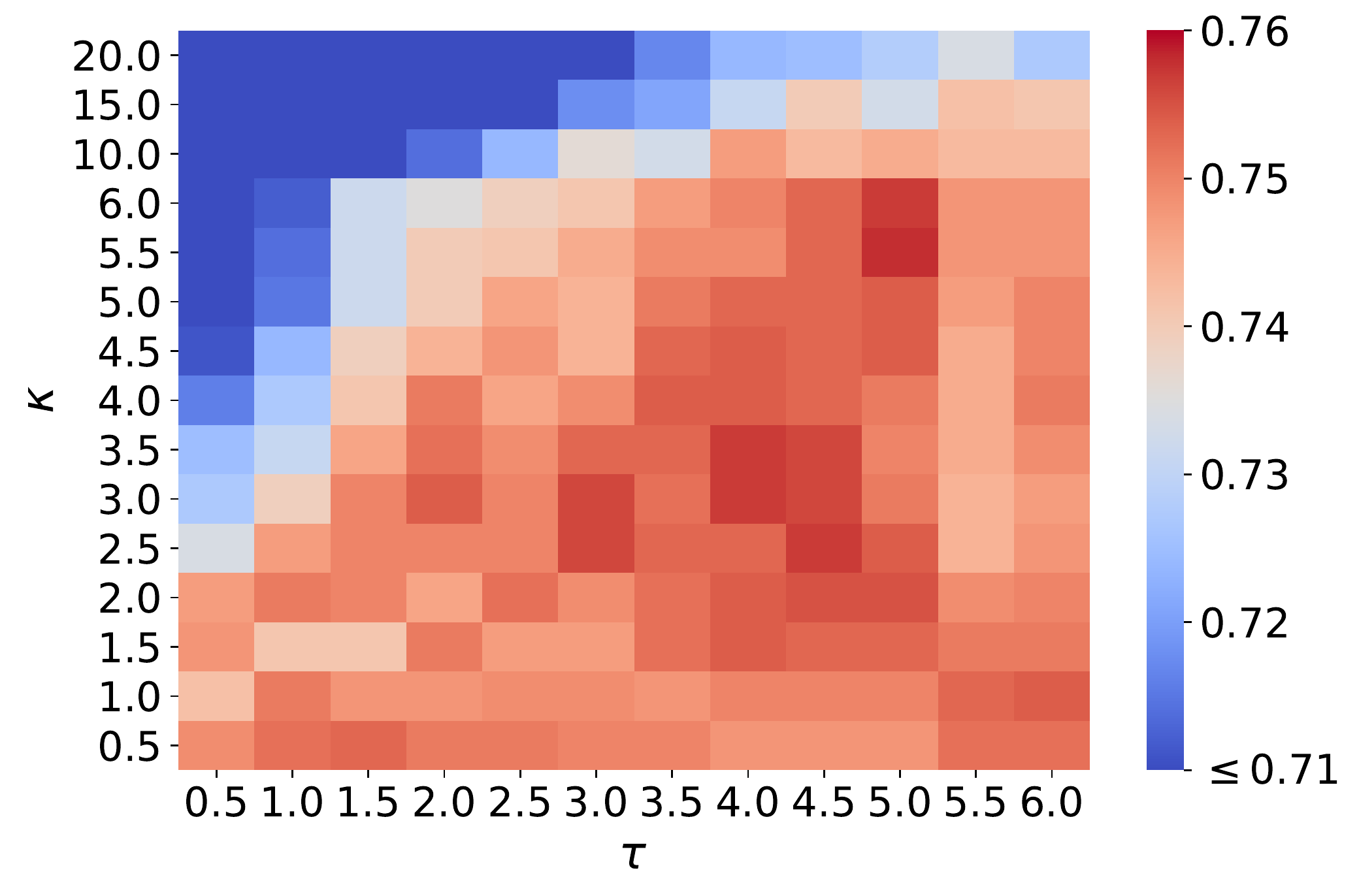}
		\caption{Mean curvature (Lorentz kernel)} 
	\end{subfigure}
	\caption{Pearson correlation coefficients  $(R_p)$ from 5-fold cross-validation of ${\rm EIC}^{H}_{{\alpha}, \kappa,\tau}$ on the PDBbind v2015 refined set (excluding the core set) are plotted against different values of $\tau$ and $\kappa$. The element interactive  mean curvatures are utilized for all calculations.
	The best parameters and median values of $R_p$ for  ${\rm EIC}^{H}_{{\alpha}, \kappa,\tau}$ are respectively found to be 
		(a) exponential-kernel model: $(\tau=5.0, \kappa=1.5, R_p=0.754)$;
		(b) Lorentz-kernel model: $(\tau=5.0, \kappa=5.5, R_p=0.758)$
	}
	\label{fig:v2015-el-kfold}
\end{figure}

Figure \ref{fig:v2007-eell-kfold} presents the 5-fold CV performances of ${\rm EIC}^{HH}_{{\alpha}, \kappa_1,\tau_1; {\alpha}, \kappa_2,\tau_2}$ on  the training set of the PDBbind v2007 benchmark against the different choices of $\kappa_2$ and $\tau_2$. Parameters for the first kernel $\kappa_1$ and $\tau_1$ are chosen from the report in Fig. \ref{fig:v2007-el-kfold}
Figure \ref{fig:v2015-el-kfold} depicts the 5-fold CV performance of ${\rm EIC}^{H}_{{\alpha}, \kappa,\tau}$ on the training set of the PDBbind v2013 benchmark against the different choices of $\kappa$ and $\tau$.

\begin{figure}
	\centering
	\begin{subfigure}{0.4\textwidth} 
		\includegraphics[width=\textwidth]{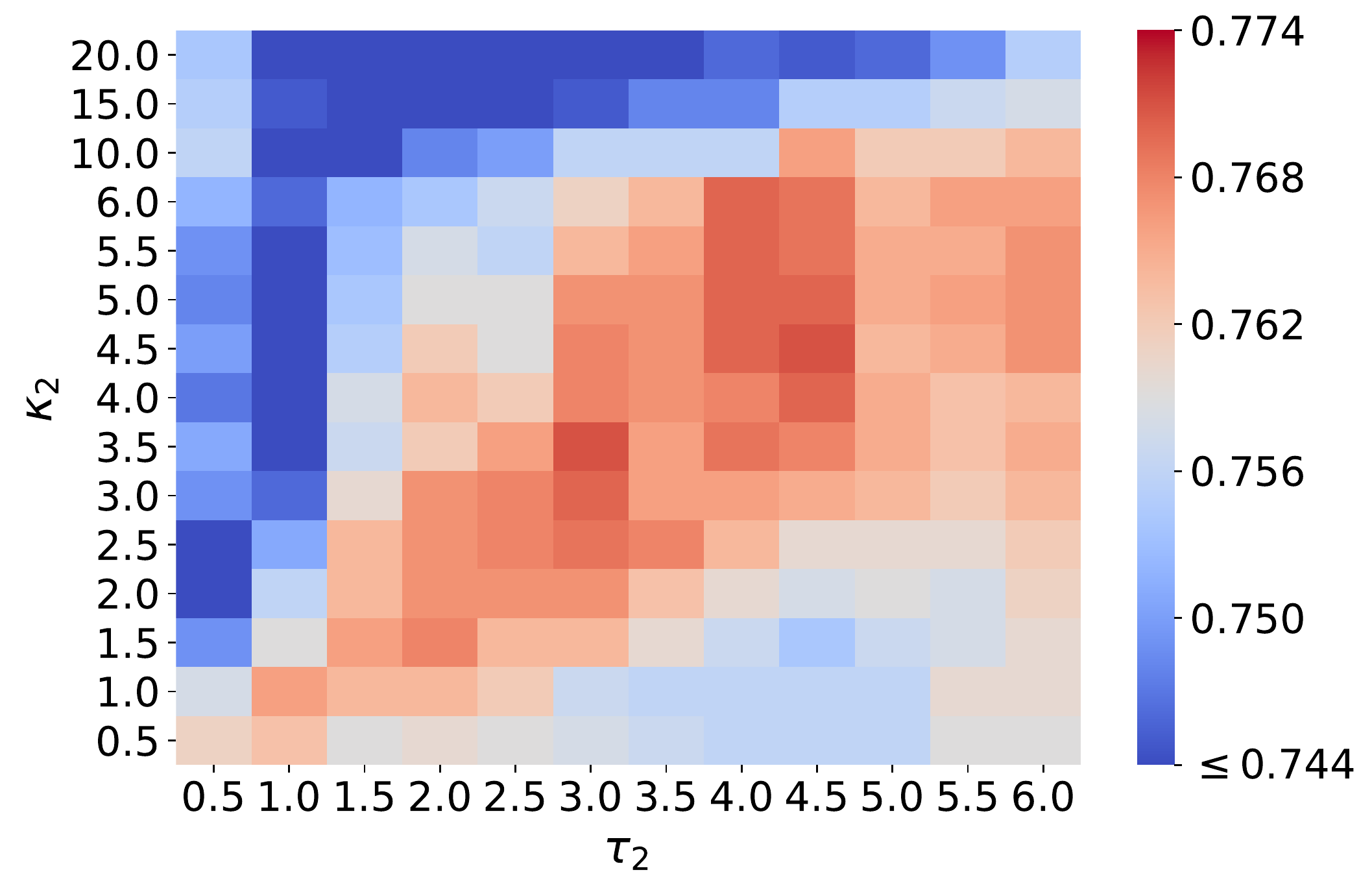}
		\caption{Mean curvature (exponential kernel)} 
	\end{subfigure}
	\vspace{1em} 
	\begin{subfigure}{0.4\textwidth} 
		\includegraphics[width=\textwidth]{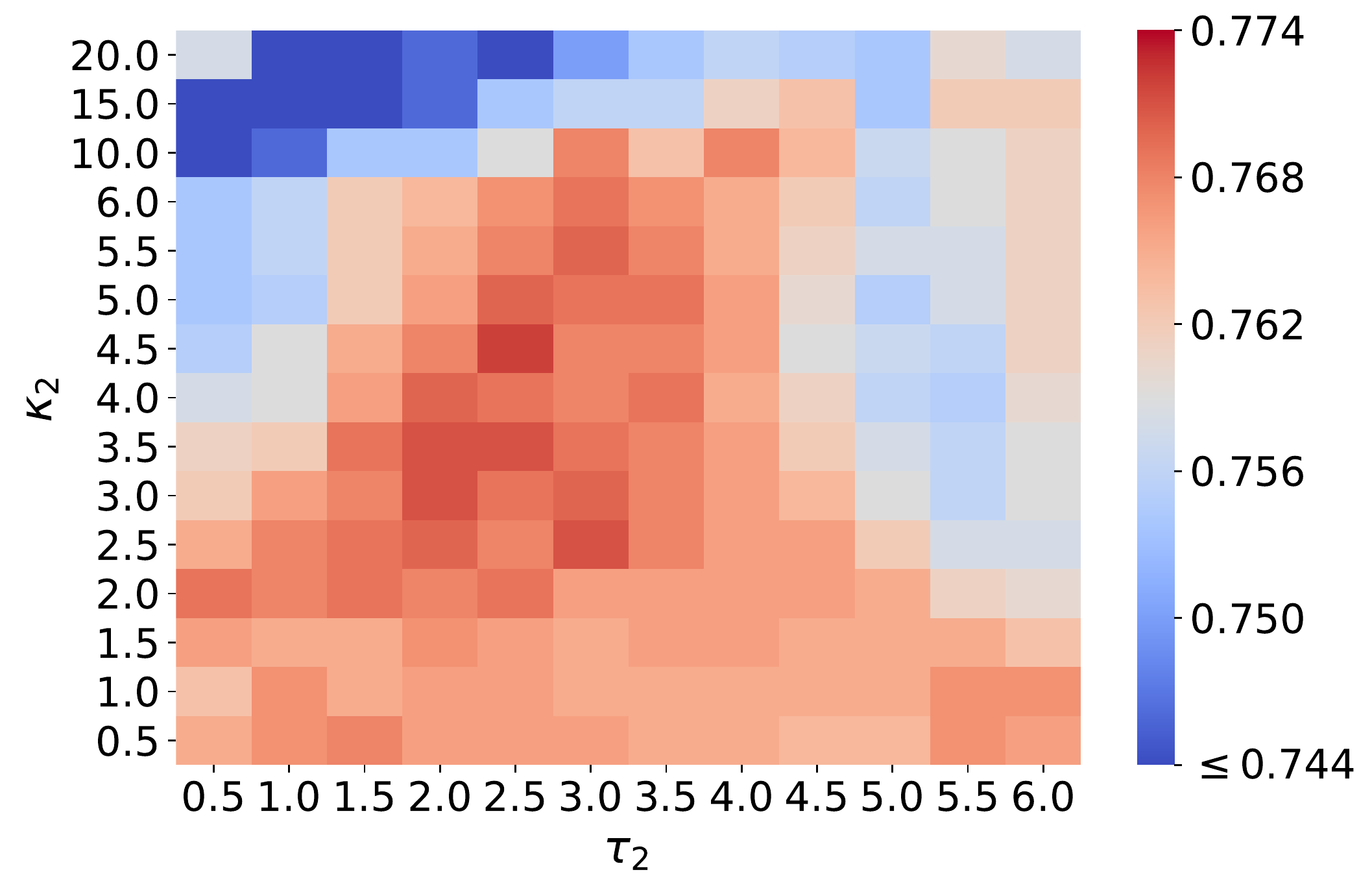}
		\caption{Mean curvature (Lorentz kernel)} 
	\end{subfigure}
	\caption{Pearson correlation coefficients  $(R_p)$ from 5-fold cross-validation of ${\rm EIC}^{HH}_{{\alpha}, \kappa_1,\tau_1; {\alpha}, \kappa_2,\tau_2}$ on the PDBbind v2015 refined set (excluding the core set) are plotted against different values of $\tau_2$ and $\kappa_2$. The element interactive  mean curvatures are utilized for all calculations. While the parameters of the first kernel $(\tau_1, \kappa_1)$ are fixed, the parameters of the second kernel  $(\tau_2, \kappa_2)$ are varied in the interested domains. The best parameters and median values of $R_p$ for ${\rm EIC}^{HH}_{{\alpha}, \kappa_1,\tau_1; {\alpha}, \kappa_2,\tau_2}$ are respectively found to be 
		(a) exponential-kernel model: $(\tau_1=5.0, \kappa_1=1.5, \tau_2=3.0, \kappa_2=3.5, R_p=0.771)$;
		(b) Lorentz-kernel model: $(\tau_1=5.0, \kappa_1=5.5, \tau_2=2.5, \kappa_2=4.5, R_p=0.772)$
	}
	\label{fig:v2015-eell-kfold}
\end{figure}

Figure \ref{fig:v2015-eell-kfold} reveals the 5-fold CV performance of ${\rm EIC}^{HH}_{{\alpha}, \kappa_1,\tau_1; {\alpha}, \kappa_2,\tau_2}$ on the training set of the PDBbind v2015 benchmark against the different choices of $\kappa_2$ and $\tau_2$. Parameters for the first kernel $\kappa_1$ and $\tau_1$ are chosen from those  reported in Fig. \ref{fig:v2015-el-kfold}

\subsubsection{Parameter search using the test set}
Table \ref{tab:PDBbindv2007_results_testset} lists the performances of EIC models with kernel parameters optimized based on PDBbind v2007 core set validation.
\begin{table*}[!ht]
	\centering
	\caption{Predictive performance of various models on PDBbind v2007 benchmark. Here, notation $^*$ indicates the EIC kernel parameters were optimized according to the test set performance.}
	\begin{tabular}{|c|c|c|}
		\hline
		Method & $R_p$ & RMSE (kcal/mol) \\\hline
		$^*$EIC$^{H}_{{\rm E},2,1}$   & 0.802 & 2.069 \\
		$^*$EIC$^{HH}_{{\rm E},2,1;{\rm E},4,4.5}$  & 0.815 & 2.003 \\
		$^*$EIC$^{H}_{{\rm L},2,0.5}$  & 0.797 & 2.048 \\
		$^*$EIC$^{HH}_{{\rm L},2,0.5;{\rm L},10,4}$   & 0.819 & 1.968 \\
		$^*$Consensus$^H$     & {0.825} & {1.967} \\
		\hline 
	\end{tabular}
	\label{tab:PDBbindv2007_results_testset}
\end{table*}

Table \ref{tab:PDBbindv2015_results_testset} lists the performances of EIC models with kernel parameters optimized based on the PDBbind v2013 core set validation.
\begin{table*}[!ht]
	\centering
	\caption{Predictive performance of various models on the PDBbind v2013 benchmark. Here, notation $^*$ indicates the EIC kernel parameters were optimized according to the test set performance.}
	\begin{tabular}{|c|c|c|}
		\hline
		Method & $R_p$ & RMSE (kcal/mol) \\
		\hline
		$^*$EIC$^{H}_{E,0.5,5.5}$    & 0.780 & 2.001 \\
		$^*$EIC$^{HH}_{E,0.5,5.5;E,6,5}$   & 0.790 & 1.972 \\
		$^*$EIC$^{H}_{L,1,5.5}$  & 0.784 & 1.993 \\
		$^*$EIC$^{HH}_{L,1,5.5;L,6,6}$   & {0.798} & {1.951} \\
		$^*$Consensus$^H$     & {0.798} & 1.952 \\
		\hline 
	\end{tabular}
	\label{tab:PDBbindv2015_results_testset}
\end{table*}

Table \ref{tab:PDBbindv2016_results_testset} lists the performances of EIC models with kernel parameters optimized based on PDBbind v2016 core set validation.
\begin{table*}[!ht]
	\centering
	\caption{Predictive performance of various models on the PDBbind v2016 benchmark. Here, notation $^*$ indicates the EIC kernel parameters were optimized according to the test set performance.}
	\begin{tabular}{|c|c|c|}
		\hline
		Method & $R_p$ & RMSE (kcal/mol) \\
		\hline
		$^*$EIC$^{H}_{E,0.5,5.5}$    & 0.819 & 1.801 \\
		$^*$EIC$^{HH}_{E,0.5,5.5;E,6,5}$   & 0.827 & 1.763 \\
		$^*$EIC$^{H}_{L,1,5.5}$  & 0.824 & 1.785 \\
		$^*$EIC$^{HH}_{L,1,5.5;L,6,6}$   & 0.831 & 1.754 \\
		$^*$Consensus$^H$     & {0.834} & {1.746} \\
		\hline 
	\end{tabular}
	\label{tab:PDBbindv2016_results_testset}
\end{table*}


\end{document}